\DocumentMetadata{}
\documentclass[manuscript,nonacm]{acmart}
\usepackage{balance}       
\usepackage{graphics}      
\usepackage{hyperref}
\usepackage{color}
\usepackage{booktabs}
\usepackage{textcomp}
\usepackage{enumerate}
\usepackage{lipsum}
\usepackage{makecell}
\usepackage{multicol}
\usepackage{multirow}
\usepackage{array}
\usepackage{verbatimbox}
\usepackage{enumitem}
\usepackage{amsmath}
\usepackage{stfloats}
\usepackage{amsthm}
\usepackage{listings}
\usepackage{caption} 
\usepackage{xspace}
\usepackage[linesnumbered]{algorithm2e}
\usepackage{algpseudocode}
\usepackage{tabularx}
\usepackage[bottom]{footmisc}
\usepackage{stackengine}
\usepackage{placeins}
\usepackage{graphicx}
\usepackage{subcaption} 
\usepackage{pifont} 
\usepackage{longtable}

\newcolumntype{L}[1]{>{\raggedright\let\newline\\\arraybackslash\hspace{0pt}}m{#1}}
\newcolumntype{C}[1]{>{\centering\let\newline\\\arraybackslash\hspace{0pt}}m{#1}}
\newcolumntype{R}[1]{>{\raggedleft\let\newline\\\arraybackslash\hspace{0pt}}m{#1}}

\makeatletter
\def\thickhline{%
  \noalign{\ifnum0=`}\fi\hrule \@height \thickarrayrulewidth \futurelet
   \reserved@a\@xthickhline}
\def\@xthickhline{\ifx\reserved@a\thickhline
               \vskip\doublerulesep
               \vskip-\thickarrayrulewidth
             \fi
      \ifnum0=`{\fi}}
\makeatother

\makeatletter
\def\thickhlinespace{%
  \addlinespace[1ex]
  \noalign{\ifnum0=`}\fi\hrule \@height \thickarrayrulewidth \futurelet
   \reserved@a\@xthickhline
   \addlinespace[1ex]
   }
\def\@xthickhlinespace{\ifx\reserved@a\thickhline
               \vskip\doublerulesep
               \vskip-\thickarrayrulewidth
             \fi
      \ifnum0=`{\fi}}
\makeatother

\newlength{\thickarrayrulewidth}
\newlength\Origarrayrulewidth

\algnewcommand{\IfThenElse}[3]{
  \State \algorithmicif\ #1\ \algorithmicthen\ #2\ \algorithmicelse\ #3}

\newcommand{\systemName}{Thought-GPS\xspace}
\newcommand{\baseline}{Active Control\xspace}
\newcommand{\bstat}[2]{%
    $\beta$=#1, 95\% CrI $[#2]$\xspace
}

\definecolor{onegreen}{HTML}{255634}
\definecolor{twored}{HTML}{B5404A}
\definecolor{threeblue}{HTML}{4675AF}
\definecolor{fourpurple}{HTML}{9661BC}

\newcommand{\dingOne}{\textcolor{onegreen}{\ding{182}}\xspace}
\newcommand{\dingTwo}{\textcolor{twored}{\ding{183}}\xspace}
\newcommand{\dingThree}{\textcolor{threeblue}{\ding{184}}\xspace}
\newcommand{\dingFour}{\textcolor{fourpurple}{\ding{185}}\xspace}

\definecolor{downredcolor}{HTML}{e31a1c}
\definecolor{upgreencolor}{HTML}{33a02c}

\definecolor{DarkGreen}{HTML}{5DAC81}

\usepackage{tikz}
\usepackage{xparse}

\definecolor{plotdata}{HTML}{1F77B4}       %
\definecolor{plotmean}{HTML}{D95F00}       %
\definecolor{plotmeanlabel}{HTML}{994D00}  %
\definecolor{plotticks}{HTML}{666666}      %
\newlength{\dotplotheight}
\newlength{\dotplotwidth}
\pgfmathsetlengthmacro{\plotwidthnum}{\dotplotwidth}
\pgfmathsetlengthmacro{\plotheightnum}{\dotplotheight}

\pgfmathsetmacro{\tickh}{0.2}    
\pgfmathsetmacro{\meanh}{0.5}     
\pgfmathsetmacro{\pointsize}{0.12}

\ExplSyntaxOn
\keys_define:nn { dotplot }
 {
  min  .tl_set:N  = \l_dotplot_min_tl,
  min  .initial:n = {0},
  max  .tl_set:N  = \l_dotplot_max_tl,
  max  .initial:n = {10},
  step .tl_set:N  = \l_dotplot_step_tl,
  step .initial:n = {1},
 }
\ExplSyntaxOff

\ExplSyntaxOn
\NewDocumentCommand{\dotplot}{ O{} m }
 {
  \group_begin:
  \keys_set:nn { dotplot } { #1 }

  \pgfmathsetmacro{\dotplotSum}{0}
  \pgfmathsetmacro{\dotplotCount}{0}
  \foreach \val in {#2} {
      \pgfmathparse{\dotplotSum + \val} \global\let\dotplotSum\pgfmathresult
      \pgfmathparse{\dotplotCount + 1} \global\let\dotplotCount\pgfmathresult
  }
  \pgfmathsetmacro{\mean}{ifthenelse(\dotplotCount>0,\dotplotSum/\dotplotCount,0)}

  \pgfmathsetmacro{\numticks}{int((\l_dotplot_max_tl - \l_dotplot_min_tl)/\l_dotplot_step_tl)}

\begin{tikzpicture}[x=\plotwidthnum/(\l_dotplot_max_tl - \l_dotplot_min_tl), y=\plotheightnum, baseline=-0.25\dotplotheight]
    \draw[plotticks] (\l_dotplot_min_tl,0) -- (\l_dotplot_max_tl,0);

    \pgfmathsetlengthmacro{\tickheightnum}{\tickh*\dotplotheight}
    \foreach \i in {0,...,\numticks} {
        \pgfmathsetmacro{\x}{\l_dotplot_min_tl + \i*\l_dotplot_step_tl}
        \ifnum\i=0
            \draw[plotticks, thick] (\x,-\tickheightnum) -- (\x,\tickheightnum);
            \node[overlay, font=\tiny, anchor=north, gray] at (\x+0.15,\tickheightnum-1) {\l_dotplot_min_tl};
        \else
            \ifnum\i=\numticks
                \draw[plotticks, thick] (\x,-\tickheightnum) -- (\x,\tickheightnum);
                \node[overlay, font=\tiny, anchor=north, gray] at (\x-0.15,\tickheightnum-1) {\l_dotplot_max_tl};
            \else
                \draw[plotticks] (\x,0) -- (\x,\tickheightnum);
            \fi
        \fi
    }

    \pgfmathsetlengthmacro{\radiusnum}{\pointsize*\dotplotheight}
    \def\jitter{0.15} %
    
    \foreach \val in {#2} {
      \pgfmathsetmacro{\yjitter}{\jitter*rand}
      \fill[plotdata, opacity = 0.6] (\val,\yjitter) circle[radius=\radiusnum];
    }

    \pgfmathsetlengthmacro{\meanheightnum}{\meanh*\dotplotheight}
    \draw[plotmean, line~width=1pt] (\mean,-\meanheightnum) -- (\mean,\meanheightnum);
    \node[overlay, font={\bfseries\tiny}, anchor=west, gray, plotmean] 
    at (\mean-0.1, \tickheightnum+2.2) 
    {\pgfmathprintnumber[fixed, precision=2]{\mean}};
  \end{tikzpicture}

  \group_end:
 }
\ExplSyntaxOff

\ExplSyntaxOn
\NewDocumentCommand{\pieplot}{ O{plotticks} m m }
 {
  \pgfmathsetmacro{\pieangle}{360*(#2)/(#3)}
  
  \begin{tikzpicture}[baseline=-0.8ex]
    \draw[#1] (0,0) circle (0.13cm);
    \fill[#1!60] (0,0) -- (0,0.13cm)
      arc [start~angle=90, end~angle={90-\pieangle}, radius=0.13cm] -- cycle;
  \end{tikzpicture}%
 }
\ExplSyntaxOff

\newcommand{\progressplot}[4][3.0]{%
  \begin{tikzpicture}[baseline=-0.25ex]
    \pgfmathsetmacro{\percentage}{(#2 - #3)/(#4 - #3)}
    \pgfmathsetmacro{\barwidth}{#1}
    \pgfmathsetmacro{\progresswidth}{\percentage * \barwidth}
    
    \fill[color=lightgray!50] (0, 0) rectangle +(\barwidth ex, 1.2ex);
    
    \fill[color=plotdata!75] (0, 0) rectangle +(\progresswidth ex, 1.2ex);

    \node[font=\bfseries\small, anchor=west, text=black, inner sep=1pt] 
      at (\barwidth ex + 0.3ex, 0.6ex) 
      {\pgfmathprintnumber[fixed, precision=2]{#2}};
  \end{tikzpicture}%
}

\definecolor{maroon}{cmyk}{0, 0.87, 0.68, 0.32}
\definecolor{halfgray}{gray}{0.55}
\definecolor{ipython_frame}{RGB}{207, 207, 207}
\definecolor{ipython_bg}{RGB}{247, 247, 247}
\definecolor{ipython_red}{RGB}{186, 33, 33}
\definecolor{ipython_green}{RGB}{0, 128, 0}
\definecolor{ipython_cyan}{RGB}{64, 128, 128}
\definecolor{ipython_purple}{RGB}{170, 34, 255}
\lstdefinelanguage{Markdown}{
    basicstyle=\ttfamily\footnotesize,
}
\lstdefinelanguage{iPython}{
    morekeywords={access,and,break,class,continue,def,del,elif,else,except,exec,finally,for,from,global,if,import,in,is,lambda,not,or,pass,print,raise,return,try,while},%
    morekeywords=[2]{abs,all,any,basestring,bin,bool,bytearray,callable,chr,classmethod,cmp,compile,complex,delattr,dict,dir,divmod,enumerate,eval,execfile,file,filter,float,format,frozenset,getattr,globals,hasattr,hash,help,hex,id,input,int,isinstance,issubclass,iter,len,list,locals,long,map,max,memoryview,min,next,object,oct,open,ord,pow,property,range,raw_input,reduce,reload,repr,reversed,round,set,setattr,slice,sorted,staticmethod,str,sum,super,tuple,type,unichr,unicode,vars,xrange,zip,apply,buffer,coerce,intern},%
    sensitive=true,%
    morecomment=[l]\#,%
    morestring=[b]',%
    morestring=[b]",%
    morestring=[s]{'''}{'''},
    morestring=[s]{"""}{"""},
    morestring=[s]{r'}{'},
    morestring=[s]{r"}{"},%
    morestring=[s]{r'''}{'''},%
    morestring=[s]{r"""}{"""},%
    morestring=[s]{u'}{'},
    morestring=[s]{u"}{"},%
    morestring=[s]{u'''}{'''},%
    morestring=[s]{u"""}{"""},%
    literate=
    {á}{{\'a}}1 {é}{{\'e}}1 {í}{{\'i}}1 {ó}{{\'o}}1 {ú}{{\'u}}1
    {Á}{{\'A}}1 {É}{{\'E}}1 {Í}{{\'I}}1 {Ó}{{\'O}}1 {Ú}{{\'U}}1
    {à}{{\`a}}1 {è}{{\`e}}1 {ì}{{\`i}}1 {ò}{{\`o}}1 {ù}{{\`u}}1
    {À}{{\`A}}1 {È}{{\'E}}1 {Ì}{{\`I}}1 {Ò}{{\`O}}1 {Ù}{{\`U}}1
    {ä}{{\"a}}1 {ë}{{\"e}}1 {ï}{{\"i}}1 {ö}{{\"o}}1 {ü}{{\"u}}1
    {Ä}{{\"A}}1 {Ë}{{\"E}}1 {Ï}{{\"I}}1 {Ö}{{\"O}}1 {Ü}{{\"U}}1
    {â}{{\^a}}1 {ê}{{\^e}}1 {î}{{\^i}}1 {ô}{{\^o}}1 {û}{{\^u}}1
    {Â}{{\^A}}1 {Ê}{{\^E}}1 {Î}{{\^I}}1 {Ô}{{\^O}}1 {Û}{{\^U}}1
    {œ}{{\oe}}1 {Œ}{{\OE}}1 {æ}{{\ae}}1 {Æ}{{\AE}}1 {ß}{{\ss}}1
    {ç}{{\c c}}1 {Ç}{{\c C}}1 {ø}{{\o}}1 {å}{{\r a}}1 {Å}{{\r A}}1
    {€}{{\EUR}}1 {£}{{\pounds}}1
    {^}{{{\color{ipython_purple}\^{}}}}1
    {=}{{{\color{ipython_purple}=}}}1
    {+}{{{\color{ipython_purple}+}}}1
    {*}{{{\color{ipython_purple}$^\ast$}}}1
    {/}{{{\color{ipython_purple}/}}}1
    {+=}{{{+=}}}1
    {-=}{{{-=}}}1
    {*=}{{{$^\ast$=}}}1
    {/=}{{{/=}}}1,
    literate=
    *{-}{{{\color{ipython_purple}-}}}1
     {?}{{{\color{ipython_purple}?}}}1,
    identifierstyle=\color{black}\ttfamily,
    commentstyle=\color{ipython_cyan}\ttfamily,
    stringstyle=\color{ipython_red}\ttfamily,
    keepspaces=true,
    showspaces=false,
    showstringspaces=false,
    basicstyle=\ttfamily\footnotesize,
    keywordstyle=\color{ipython_green}\ttfamily,
}

\setcopyright{none}
\acmDOI{}
\acmISBN{}
\acmPrice{}

\begin{document}
\pagestyle{plain}

%

\title{``\textit{It Wouldn't Have Been So Apparent to Me}'': Evaluating How Thought-Level Progress Tracking Supports Psychosocial Goals}

%

\author{Shiyu Xu}
\email{shiyuxu@umich.edu}
\affiliation{%
  \institution{University of Michigan}
  \city{Ann Arbor}
  \state{Michigan}
  \country{United States}
}

\author{Ruishi Zou}
\email{ruishizou@g.harvard.edu}
\affiliation{%
  \institution{Harvard University}
  \city{Cambridge}
  \state{Massachusetts}
  \country{United States}
}

\author{Anna R. Van Meter}
\email{anna.vanmeter@nyulangone.org}
\affiliation{%
  \department{Child and Adolescent Psychiatry}
  \institution{NYU Grossman School of Medicine}
  \city{New York}
  \state{New York}
  \country{United States}
}

\author{Addie Weaver}
\email{weaverad@umich.edu}
\affiliation{%
  \department{School of Social Work}
  \institution{University of Michigan}
  \city{Ann Arbor}
  \state{Michigan}
  \country{United States}
}

\author{Emily Mower Provost}
\email{emilykmp@umich.edu}
\affiliation{%
  \department{Computer Science and Engineering}
  \institution{University of Michigan}
  \city{Ann Arbor}
  \state{Michigan}
  \country{United States}
}

\author{Daniel A. Adler}
\email{adlerdan@umich.edu}
\affiliation{%
  \department{Computer Science and Engineering}
  \institution{University of Michigan}
  \city{Ann Arbor}
  \state{Michigan}
  \country{United States}
}

%
\renewcommand{\shortauthors}{Xu et al.}
\renewcommand{\shorttitle}{}

%
\begin{abstract}
We pursue psychosocial goals every day, from coping with anxiety, to pursuing personal growth. While prior research has designed systems for tracking progress towards behavioral goals, fewer studies have focused on progress towards psychosocial goals -- and the thinking changes people make to address them. As a first step, we present \systemName, a system that visualizes potential markers of such thought-level progress in-the-moment as users interact with an AI-supported cognitive restructuring intervention. We evaluated \systemName through a mixed-methods single-session study with 30 participants, comparing \systemName with an active control condition without progress markers. We found preliminary evidence that participants using \systemName reported greater clarity about how they might change to pursue their goals, stronger motivation to reflect on goal progress, and greater engagement in that reflection. Participants also reported that progress markers made subtle progress visible, helped them engage in deeper reflection, and connect conversational insights to their personal goals.
\end{abstract}

%
%
%

\begin{teaserfigure}
    \centering
    \includegraphics[width=1\linewidth]{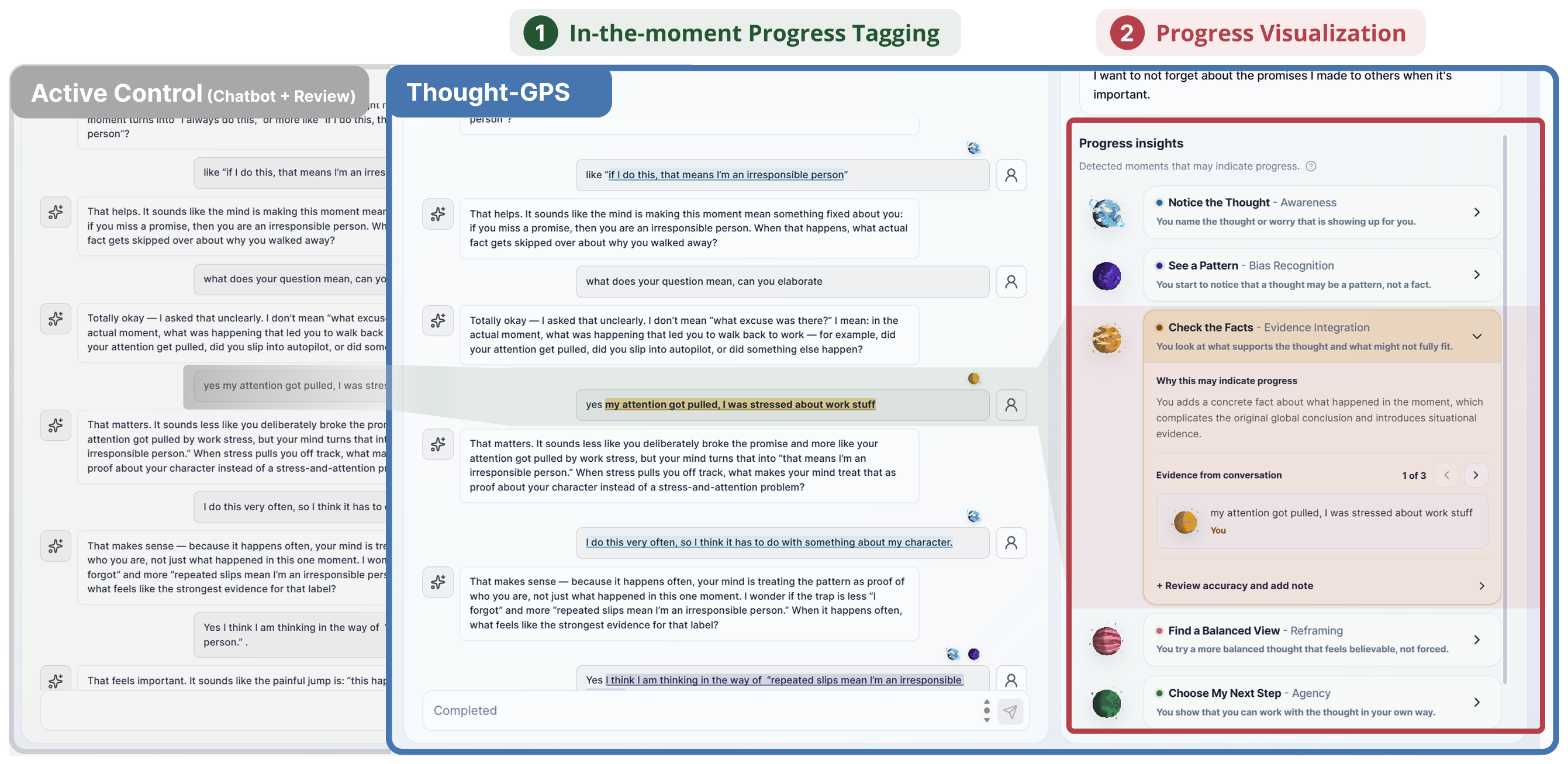}
    \caption{Overview of our proposed system \systemName. Alongside a psychotherapeutic chatbot practicing cognitive restructuring for psychosocial goals, the system supports \dingOne in-the-moment progress tagging and \dingTwo progress visualization. We hypothesize that this set of in-the-moment conversation-progress visualizations can help users notice and interpret changes in their thinking during the conversation.}
    \label{fig:teaser}
    \Description{}
\end{teaserfigure}

%
\maketitle

\section{Introduction}
\label{sec:introduction}

People pursue many kinds of personal goals throughout their lives. Some goals target physical or behavioral outcomes, such as exercising more regularly or completing planned tasks~\cite{consolvo2008activity, jorke2026bloom}. Others concern cognitive, emotional, social, or relational changes, such as responding differently to anxious thoughts or communicating more openly in a relationship~\cite{clark2013cognitive, agapie2022longitudinal}. In practice, however, these dimensions often overlap: pursuing a goal may require changes in behavior, changes in how people interpret and respond to their experiences, or both~\cite{sas2025beyond}.

A body of work in human-computer interaction has explored the design of technologies that help people track progress toward their personal goals~\cite{ekhtiar2023goals}. Much of this work has focused on tracking markers of progress involving behavioral changes that can be explicitly recorded or observed, such as the number of steps taken, exercise sessions completed, or tasks finished~\cite{consolvo2008activity, bentley2013health, lee2024goaltrack, jorke2026bloom}. We refer to progress reflected in such observable behavioral changes as \textit{behavior-level progress}. By making such changes visible over time, these systems support self-awareness, foster motivation, and facilitate change~\cite{li2010stage, epstein2015lived, zhu2025systematic}.

Despite the demonstrated value of progress tracking, there has been less focus on how technology can help people recognize and interpret progress towards personal goals to improve one's psychological experiences, relationships, and functioning---what we call in this work \textit{psychosocial goals} \cite{ro2010conceptualization, lam2011clinical, slovak2024hci}. Progress toward these goals may first appear through changes in how people think, particularly in how they understand and respond to their experiences, before becoming visible through behavior or other measurable outcomes (e.g., symptom, well-being, or functioning scales)~\cite{stiles1990assimilation, muran1995linking, gomezpenedo2020theory}. 
We refer to such changes in thinking as \textit{thought-level progress}. 
People often work toward such thought-level progress through structured forms of reflection~\cite{beck1979cognitive,clark2013cognitive,rime2009emotion}
using thought records or journals~\cite{kazantzis2016quantity}. However, recording an experience does not necessarily clarify or show evidence of how a person's thinking has changed, why that change might represent progress toward their psychosocial goals~\cite{lorenzoluaces2015complicated}, and existing technologies do not support such sensemaking of thought-level progress. 
Motivated by this gap, we ask: \textbf{\textit{How can we design a system to help people recognize and make sense of thought-level progress towards psychosocial goals?}}

Towards a first study to address this question, in this work, we explore how users respond to markers of thought-level progress within a conversational AI intervention that supports users on addressing barriers towards attaining their psychosocial goals~\cite{sharma2024facilitating,fitzpatrick2017delivering,heinz2025randomized}. Specifically, we designed \systemName, a web-based, goal-centered AI system for supporting progress sensemaking during a single guided conversation, in order to understand how in-the-moment conversational markers of thought-level progress shape perceived progress and reflection. 
\systemName begins with end-users defining a psychosocial goal they wish to work on, and guides users through an AI-supported conversation using the evidence-based practice of cognitive restructuring~\cite{clark2013cognitive,sharma2024facilitating}. 
The system identifies five theory-informed categories of thinking changes~\cite{clark2013cognitive} within the conversational text, which are displayed in Progress Scaffold Cards in real time during the conversation and are linked bidirectionally to highlighted excerpts from the user's own statements. Users can inspect the interpretations, review the supporting evidence, indicate whether an interpretation feels accurate, and review a category-organized summary after the conversation.

We evaluated \systemName through a mixed-method between-subjects study with 30 participants. To understand the effect of our design, we compared \systemName with an active control condition, in which users received the conversational AI intervention,  but did not see the real-time progress markers. We observed that \systemName helped participants notice and make sense of thought-level progress that might otherwise have gone unnoticed, and that participants using \systemName reported greater clarity and reflective engagement relative to the active control condition, with comparable usability across conditions, measured using adapted questions from standardized assessments of working alliance, technology-supported reflection, and self-efficacy towards goal attainment. At the same time, participants in both conditions differentiated within-conversation progress from more observable markers of goal attainment, and diverged in when they wanted progress markers delivered, valuing real-time cues for their salience and retrospective summaries for synthesis.

In summary, our main contributions are as follows:
\vspace{-2px}
\begin{itemize}
    \item The \systemName system as a design prototype, which implements in-the-moment, AI-identified progress markers on top of a cognitive restructuring chatbot intervention that supports people’s understanding of progress toward their psychosocial goals.
    \item Findings from a randomized between-subjects study with (N=30) participants comparing \systemName with an active control condition. We analyzed quantitative survey results and in-system pre–post affect ratings alongside qualitative coding of semi-structured interviews.
    \item Design implications derived from our findings for future technologies supporting goal tracking and progress sensemaking for psychosocial goals.
\end{itemize}

\section{Related Work}
\label{sec:related_work}

Our system is motivated by recent work in goal tracking technologies, psychosocial goal setting, and mental health support chatbots. 
We first describe recent work studying goal tracking systems to provide an overview of the existing technologies that help people track and make sense of their goals.
We then focus on psychosocial goals, how they are traditionally tracked, and how people make progress towards attaining them. 
Finally, we give an overview of conversational AI mental health chatbots---our specific context of application---describing how the conversational material produced by these chatbots can be reused and interpreted to support psychosocial goal attainment. 

\subsection{Systems for Tracking and Making Sense of Goals}
\label{subsec:related_work:progress_sensemaking}

Previous research in HCI has examined how technology can help people track and reflect on progress toward their goals~\cite{li2010stage,baumer2015reflective,epstein2015lived}. Much of this work has focused on behavioral goals, in which progress can be represented through observable activities or outcomes~\cite{ekhtiar2023goals,zhu2025systematic,lee2024goaltrack}. For example, the UbiFit Garden mapped sensed physical activity onto a garden that grew as users exercised, directly representing recorded behavior as progress toward an activity goal, which improved participants' motivation for change ~\cite{consolvo2008activity}. Bloom extended this approach by combining a Large Language Model (LLM) health coach with activity planning, wearable tracking, and progress visualization. In the Bloom system, goal progress was represented primarily through planned and completed behaviors~\cite{jorke2026bloom}. By making behavioral progress visible over time, these systems have shown how tracking toward goals can support self-awareness, motivation, and continued behavior change~\cite{baumer2015reflective,consolvo2008activity}. Such representations, however, depend on activity that can be sensed or logged \cite{sas2025beyond}.

In addition to tracking behavioral goals, HCI researchers have studied technologies that support sensemaking and reflection on the progress of personal thoughts through lived experiences \cite{baumer2014reviewing,bentvelzen2022revisiting}. Sensemaking involves developing representations that help people organize and interpret information to answer questions relevant to specific tasks~\cite{russell1993cost,pirolli2005sensemaking}. In the context of making sense of personal thoughts, this can involve connecting recorded events, thoughts, and feelings to understand a concern or consider possible responses \cite{cocskun2023data,haimson2023uncovering}. For example, MindfulDiary uses LLM-guided conversations to help psychiatric patients document daily experiences and revisit conversational summaries~\cite{kim2024mindfuldiary}, while MindScape combines behavioral sensing data with LLMs to generate context-aware journaling prompts and personalized insights~\cite{nepal2024mindscape}. ExploreSelf supports user-directed reflection on personal challenges through adaptive questions and iteratively updated summaries~\cite{song2025exploreself}. Together, these systems show how conversational and personal data can be used to support reflection and clarify possible goals and actions \cite{rashik2025enabled}.

The systems above surface reflective material after an experience has been recorded. A separate line of recent work instead makes goal-related information visible while an interaction is still unfolding. OnGoal tracks users' conversational goals and evaluates how LLM responses address them, providing real-time visual feedback with explanations and supporting excerpts~\cite{coscia2025ongoal}. Conversation Progress Guide visualizes subtask completion during conversational AI use to support users' self-efficacy~\cite{jeong2025conversation}. In collaborative meetings, Chen et al.~\cite{chen2025track} explored passive visualizations and active questions to help participants clarify goals and reflect on whether the discussion remains on track. These systems make aspects of an ongoing interaction inspectable in relation to a goal, and establish that in-the-moment feedback can support goal awareness during an interaction. What these systems track, however, is not thought-level progress, but whether a model satisfied a constraint, whether a subtask is complete, or whether a discussion has stayed aligned with its goals. These are properties of the interaction itself, rather than changes in the user’s thinking.

These prior works demonstrated the value of tracking observable behavioral progress and have explored technologies for reflecting on thinking. Less attention, however, has been paid to how we can identify and visualize thought-level progress towards goals. In this work, we build on these existing systems that support sensemaking and apply it to the context in which thought-level progress may be especially important---tracking psychosocial goals.

\subsection{Understanding Progress Toward Personal, Psychosocial Goals}

Psychosocial goals concern changes shaped by the relationship between a person's psychological experiences and social context, with implications for wellbeing and everyday functioning~\cite{engel1977need, ro2010conceptualization, lam2011clinical}. Examples of these goals may include managing anxiety, responding differently to stress, or improving a relationship. People make progress toward psychosocial goals both in the clinic and outside the clinic. In clinical contexts, people collaborate with physicians to translate a concern into a personally meaningful goal~\cite{agapie2022longitudinal,cooper2023goals,wollburg2010goal,kiresuk1968goal}. Outside clinical care, people may similarly develop and pursue these goals through self-guided reflection~\cite{sheldon1999goal,rime2009emotion,schleider2021open}. Progress toward these goals can involve both changes in how people understand and respond to their experiences and changes in what they subsequently do \cite{gomezpenedo2020theory,lorenzoluaces2015complicated}. This distinction motivates tracking and visualizing changes in how people think about their experiences as a form of progress toward psychosocial goals, which can further motivate downstream actions towards psychosocial goal attainment.

One approach to supporting thought-level change is through structured psychotherapeutic practices. One prevalent practice is Cognitive Restructuring (CR), which is designed to help people identify distressing thoughts, examine the assumptions and evidence underlying them, and develop more balanced interpretations~\cite{beck1979cognitive,clark2013cognitive,oewel2024approaches}. CR has been associated with improvement in addressing thinking patterns associated with distress, symptoms of mental health disorders, and their associated impairment \cite{ezawa2023cognitive,shikatani2014impact}. People may document their CR practice using thought records, worksheets, or personal journals so that they can revisit their thoughts and experiences over time~\cite{kazantzis2016quantity,strunk2007patients}. However, maintaining a record does not necessarily help a person recognize how their thinking has changed or determine whether that change represents progress toward their goal. Prior work has found that structured guidance and reflection -- typically provided by licensed psychotherapists -- can help clients interpret their CR practice and make sense of their experiences~\cite{agapie2022longitudinal,oewel2024technology}. This embodies the idea of what we call ``progress sensemaking''~\cite{russell1993cost, pirolli2005sensemaking,cocskun2023data}, supporting individuals by helping them understand how they are making progress towards their goals and why.

Recognizing thought-level progress, therefore, depends on how reflective moments are interpreted, not simply on whether they are recorded. However, it remains unclear how a system could represent thought-level progress and how users would experience that progress if it were visualized. 
In the next section, we describe the specific context in which we situate our investigation, in particular, thought-level progress markers developed on top of a conversational AI mental health support chatbot intervention.

\subsection{Conversational AI Mental Health Support Chatbots}
\label{subsec:related_work:chatbots}

In this work, we study progress tracking in the context of conversational AI mental health support chatbots, which are increasingly being used to deliver scalable mental health interventions, and which generate conversational data that can be reinterpreted as markers of thought-level progress ~\cite{fitzpatrick2017delivering, hua2025charting}. Early mental health support chatbots relied on structured or rule-based dialogue~\cite{hua2025charting,vertsberger2022adolescents}, while more recent chatbots use large language models to support more personalized dialogue that better emulate human-like conversations~\cite{hua2025charting, heinz2025randomized}. Mental health support chatbots differentiate themselves from general-purpose conversational AI tools (e.g., ChatGPT), as they are developed to follow evidence-based therapeutic strategies~\cite{fitzpatrick2017delivering, sun2025script, kuta2026effectiveness}, including CR. Prior HCI researchers have explored how CR chatbots can provide scaffolded psychosocial support~\cite{sharma2024facilitating}. These conversations provide a relevant setting for studying progress sensemaking because they help users articulate and reconsider thinking patterns related to their psychosocial concerns.

Existing conversational AI mental health interventions can provide users with summaries of their conversations~\cite{kim2024mindfuldiary, hull2026generativea}. 
However, retrospective summaries of conversational records do not necessarily help users interpret possible markers of progress: they may not make clear how a specific interaction creates progress towards a user-defined goal, or how conversational activity contributes to its pursuit \cite{kim2024mindfuldiary,coscia2025ongoal}.
This ambiguity may make it difficult for users to assess how the system's guidance aligns with their goals and whether interactions result in progress towards attaining those goals---questions reflected in the goal and task dimensions of working alliance, a key indicator of psychotherapeutic success ~\cite{hatcher2006development,darcy2021evidence,bordin1979generalizability}. Similarly, summaries may preserve what was discussed without prompting the reinterpretation associated with reflection~\cite{bentvelzen2021development,bentvelzen2022revisiting}. However, when reflective change leaves no trace outside the conversation, users have little basis for recognizing movement toward a goal, which bears on whether they see the goal as attainable and what they understand as available next steps \cite{snyder1996development,bandura1977self,geurtzen2020patients}.

In this work, we study how tracking and visualizing specific conversational text as markers of thought-level progress within conversational AI mental health interventions could be interpreted by users as progress towards attaining their specific psychosocial goals, and evaluate their potential to improve perceived goal-task alignment~\cite{hatcher2006development}, technology-supported reflection~\cite{bentvelzen2021development}, and goal-related perceptions~\cite{schimpf2026supporting, snyder1996development, stoyanov2015mobile}.

\section{System}
\label{sec:system}

Our related work (\S\ref{sec:related_work}) describes technologies and practices that can support individuals in tracking and understanding progress towards their goals. Motivated by this work, we specifically study how to scaffold progress sensemaking towards implicit psychosocial goals -- which to the best of our knowledge, has not been the focus of any existing technology or system. 
To design a tool to help users make sense of progress towards their psychosocial goals, we draw inspiration from both goal tracking systems studied within HCI and literature in psychology and psychiatry focused on traditional methods of measuring psychosocial change~\cite{agapie2022longitudinal, kiresuk1968goal, desimone2023impact, lambert2018collecting}.
In this section, we first describe our four design goals for \systemName (\S\ref{subsec:design-goals}), which draw from this literature. We then describe a formative pilot study with six adults to receive feedback on an initial prototype of \systemName, as well as feedback we received on the system design by a clinical psychologist who coauthored this work (\S\ref{subsec:formative-pilot}). We subsequently updated the system to incorporate their feedback. We describe the system design alongside its changes relative to the formative pilot in \S\ref{subsec:system-description}.

\subsection{Design Goals}
\label{subsec:design-goals}

We outline four initial design goals for \systemName:

\begin{enumerate}
    \item[\textbf{DG1}] \textbf{Support In-the-Moment Awareness of Reflective Progress.} \systemName should enable people to notice potentially meaningful markers of progress as it emerges in conversation, rather than reflecting only after the conversation has ended~\cite{baumer2015reflective, fleck2010reflecting}.

    \item[\textbf{DG2}] \textbf{Make Goal-Relevant Reflective Progress Visible and Interpretable.} \systemName should represent qualitative forms of progress expressed through conversation. These representations should help people understand how such moments may relate to progress towards a self-defined psychosocial goal.

    \item[\textbf{DG3}] \textbf{Ground Progress Representations in Conversational Evidence.} \systemName should keep representations of progress connected to the user's own language and the context in which it appeared~\cite{shneiderman1996eyes}. Grounding an interpretation in conversational evidence should help people understand how the system arrived at marking progress and allow users to interpret progress in relation to their personal goals.

    \item[\textbf{DG4}] \textbf{Preserve Users' Interpretive Agency.} \systemName should allow users to examine LLM-detected representations of progress and decide whether a representation fits their experience~\cite{amershi2019guidelines}. Uncertainty and disagreement should remain legitimate outcomes of sensemaking.
\end{enumerate}

Additionally, being a system that influences users' psychosocial wellbeing, we intended to establish safety rules bounded by the prototype's role~\cite{moore2025expressing}. We positioned \systemName as a tool to support reflection. However, \systemName is not a system that emulates the role of a clinician or delivers psychotherapy (i.e., the system should not make diagnostic or treatment claims). 
When a user's language indicates an acute mental health risk, the system should attend to immediate safety, which should take precedence over continuing the current conversation.

\subsection{Formative Pilot}
\label{subsec:formative-pilot}

\subsubsection{User Review}
We recruited six U.S. adults through Prolific. 
During the recruitment process, potential participants responded to a screening survey that assessed eligibility.
To be eligible, participants had to affirm that they were comfortable discussing a personal goal or concern with an AI system. 
Participants were between 25 and 65 years old; four identified as men and two as women. Five reported using conversational AI on most days, and one reported using it more than once a week.

Each participant completed an approximately one-hour Zoom session. After an initial interview to learn about participants' current psychosocial goal tracking practices, perceptions of progress, and prior uses of AI, participants formulated a goal and tested the initial prototype while thinking aloud.

The initial prototype highlighted moments of possible thought-level progress in the conversational text in real time and provided a post-conversation summary. The 25-minute prototype portion consisted of a 15-minute AI-guided conversation followed by an interview exploring whether the identified moments aligned with participants' perceptions of progress and their views on the summary features, feedback timing, cognitive effort, and potential improvements.

We inductively coded the interview and think-aloud data to identify recurring
patterns in participants' interpretations of the prototype, usability breakdowns, and opportunities for design refinement. These insights informed revisions to the prototype and study procedure. Because many broader patterns
were also observed in the subsequent prototype study, we synthesize converging evidence from both studies in the Findings~\ref{sec:findings} and Discussion~\ref{sec:discussion} sections while reporting pilot-specific design changes below.

\subsubsection{Expert Review}

Following the pilot, a clinical psychologist, who is also a co-author of this paper and has expertise in treating mood- and anxiety disorders with cognitive behavioral therapy, completed several practice conversations and provided feedback on the prototype. The review focused on whether the prototype is safe for participants to trial for the purposes of the study and was not intended to be a conclusive assessment of its wider safety implications, clinical efficacy, or diagnostic validity. 

Moreover, the expert provided insight into how we should design the onboarding process to help participants set their goals for the conversation. Specifically, the expert suggested using the single-session action-plan protocol to help users identify a specific psychosocial goal to work on with the CR chatbot~\cite{schleider2021open}. This informed the goal-formation flow of the final study (see \S\ref{subsec:procedure-task-study}). We document the changes made through our formative pilot, summarized in 
Table~\ref{tab:design-revisions}. 
In the following section, we describe the features of \systemName.

\begin{table*}[t]
  \centering
  \caption{Design revisions informed by the formative pilot and expert review.}
  \label{tab:design-revisions}
  \small
  \begin{tabular}{p{0.38\linewidth}p{0.15\linewidth}p{0.38\linewidth}}
    \toprule
    \textbf{Initial Design Issue} & \textbf{Feedback source} &
    \textbf{Resulting Revision} \\
     \midrule
    Broad prompts did not consistently help users identify a specific
    concern to reflect on.
    & Expert
    & Asked users to describe a difficult situation, what they wanted
    to change, and the feeling they wanted to address
    (see \textit{Goal Formation},
    \S\ref{subsubsec:system-reflection-conversation}). \\
    \midrule
    The chatbot conversation sometimes moved on before exploring the underlying
    beliefs in the user's thoughts or repeated similar questions.
    & Expert
    & Guided the conversation through defined stages, added focused
    follow-up questions and checks against repetition, and limited
    the number of turns in each stage
    (see \textit{Conversation Orchestration and Safety}
    \S\ref{subsubsec:system-conversation-safety}). \\
    \midrule

    Users unfamiliar with CBT could find the category names
    difficult to understand.
    & P1, P5; Expert
    & Added everyday labels and short descriptions alongside category
    names, such as ``Check the Facts'' for Evidence Integration.
    Expanded cards explained why an excerpt \emph{may} indicate progress
    (see \textit{Reflective Categories and Progress Scaffold Cards}
    \S\ref{subsubsec:system-progress-scaffold-cards}). \\
    \midrule

    Users sometimes missed new progress scaffold cards feedback during the conversation.
    & P1, P3, P5; Expert
    & Made card updates more noticeable, added guidance on using the
    interface, and retained the session summary for later review
    (see \textit{Reflective Categories and Progress Scaffold Cards} \S\ref{subsubsec:system-progress-scaffold-cards}
    and \textit{Retrospective Session Summary} \S\ref{subsubsec:system-progress-scaffold-cards},
    ). \\
    \midrule

    General distress and signs of an acute safety risk required
    different responses.
    & Expert
    & Continued reflection with a supportive response to general
    distress. When the model flagged an acute risk, the system paused
    the exercise and directed the user to immediate human and
    emergency support
    (see \textit{Conversation Orchestration and Safety}
    \S\ref{subsubsec:system-conversation-safety}). \\

    \bottomrule
  \end{tabular}
\end{table*}

\begin{figure}
    \centering
    \includegraphics[width=0.4\linewidth]{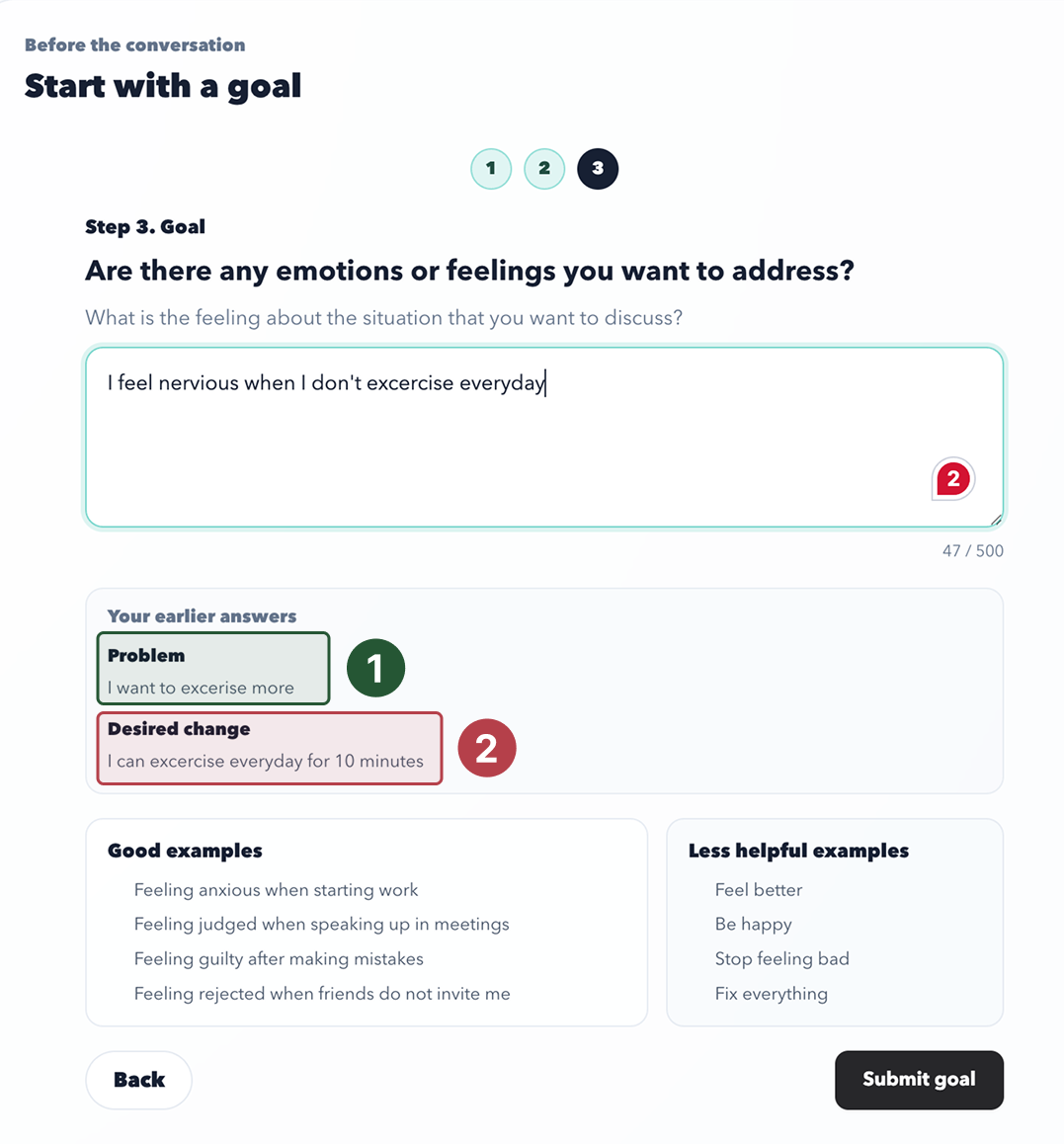}
    \caption{The goal formation process of the study. The screenshot shows the third step in goal formation, ``\textit{describe emotions or feelings.}'' The content of the two prior steps, ``\textit{describe a problem}'' \dingOne and ``\textit{describe desired change}'' \dingTwo is also present for participants to reference. }
    \label{fig:onboarding}
    \Description{}
\end{figure}

\begin{figure}
    \centering
    \includegraphics[width=1\linewidth]{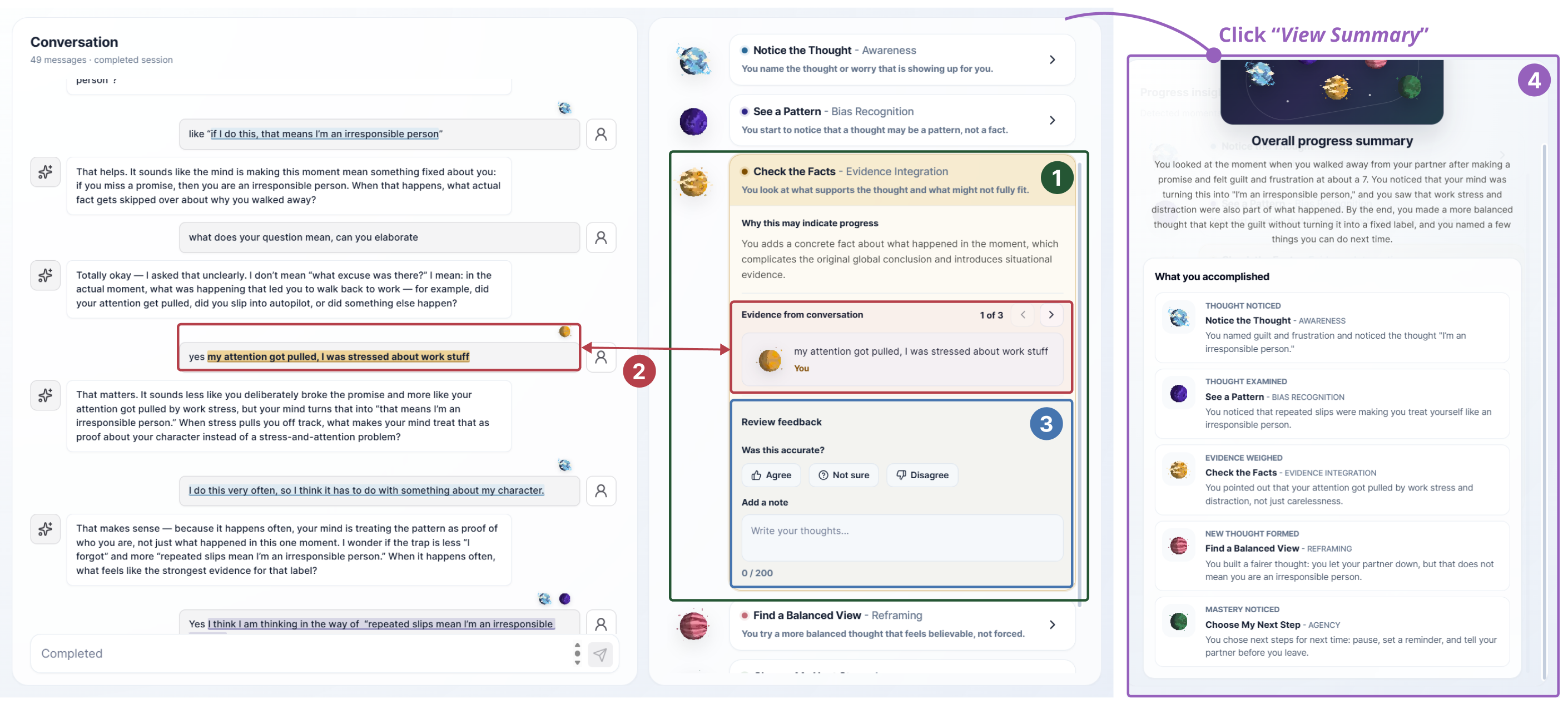}
    \caption{Overview of the \systemName interface. \systemName interacts with users through four features: progress scaffold cards \dingOne, evidence-linking mechanisms \dingTwo, goal tagging feedback \dingThree, and upon clicking a ``view summary'' button, a summary panel \dingFour that overlays on top of the progress scaffold cards.}
    \label{fig:system}
    \Description{}
\end{figure}

\subsection{\systemName}
\label{subsec:system-description}
\systemName is a web-based, LLM-supported reflection system designed to help users examine conversation activity in relation to a self-defined psychosocial goal. It combines a guided cognitive restructuring conversation with a progress-sensemaking interface that identifies and presents potentially meaningful moments from the conversation. We describe its design in the subsequent sections. 

\subsubsection{Goal-Centered Reflection Conversation}
\label{subsubsec:system-reflection-conversation}

Users began with a three-step goal-formation process that established a personally meaningful goal before entering the LLM-powered cognitive restructuring exercise.

\paragraph{Goal Formation.}
Before beginning the conversation, users completed a three-step (Fig.~\ref{fig:onboarding}) flow to establish a self-defined psychosocial goal. The flow was informed by the expert review and the single-session action-plan protocol~\cite{schleider2021open}, particularly its progression from identifying a current reason for seeking support to imagining what would be observably different if the problem improved. Users first described a recent stressful or difficult situation, then answered, ``If this problem improved, what would be different?'' Finally, they identified the emotion or feeling they wanted to address. 
This emotional concern serves in our system as the user-facing \emph{goal}, anchoring subsequent reflection and progress sensemaking in a personally meaningful concern.

\paragraph{Guided Cognitive Restructuring.}
After users completed goal formation, they navigated to the main page where they received a tutorial of the prototype. They are then invited to have a conversation with the chatbot. The prototype supported an LLM-guided conversation on cognitive restructuring. At the beginning of each session, the system incorporated the participant's self-defined goal into its opening prompt.

The conversation adapted the five-step cognitive restructuring exercise described by~\citet{sharma2024facilitating}: articulating a stressful thought, situating it in a recent context, identifying the associated emotion, examining possible thinking traps, and developing a more balanced reframe. \citet{sharma2024facilitating} elicited context through discrete form fields, presented ranked thinking-trap options, and offered generated reframes for users to select and refine. We kept this therapeutic approach but changed its form. Instead of forms and menus, ours runs as one continuous conversation. People answer in their own words and write the reframe themselves, while the model watches for relevant patterns in the background.

\paragraph{Conversation Orchestration and Safety.}
\label{subsubsec:system-conversation-safety}

The backend orchestrated the dialogue as a stage-based state machine. On each turn, the LLM received the participant's goal, the current stage, and the information accumulated so far. It returned a single structured object containing a participant-facing reply, the updated stage, a safety-risk level, and candidate detections of reflective moments. The prompt cast the model as a reflection guide rather than a clinician, so it asked focused questions in non-diagnostic language and supported participants in reaching their own interpretations \cite{sun2025script}. Deterministic rules then enforced per-stage turn limits and decided when to close the conversation.

Following expert review, we added a tiered safety policy \cite{belli2025vera,kaveladze2026risk}. General distress was met with brief care while the exercise continued. When the model assigned an acute-risk level, a deterministic guard paused restructuring, stopped the conversation from advancing, and pointed the participant to immediate human and emergency support. A rule-based fallback covered failures of the structured call \cite{kim2026pair}. These safeguards reduce foreseeable risk but do not constitute a validated risk-assessment system \cite{stamatis2026beyond}. We therefore present the prototype as a research system for LLM-supported reflection, not a diagnostic tool or clinical intervention~\cite{moore2025expressing, zhang2025computational}.
In the study itself, these system-level safeguards operated alongside live researcher supervision in every session (\S\ref{subsec:procedure-task-study}); we did not rely on the model's risk assessment as the study's safeguard.

\subsubsection{Progress-Sensemaking Interface}
\label{subsubsec:system-progress-sensemaking-interface}

While the conversation provided reflective data, the progress-sensemaking interface made selected conversational data available at three levels: category-level Progress Scaffold Cards, interpretations linked to statements from the user's own conversation, and a retrospective synthesis. These Progress Scaffold Cards and highlights were generated in real time during the conversation. We show an overview of the interface in Fig.~\ref{fig:system}.

\paragraph{Reflective Categories and Progress Scaffold Cards.}
\label{subsubsec:system-progress-scaffold-cards}
We synthesized five categories of conversational reflection, drawing on Clark's account of cognitive restructuring~\cite{clark2013cognitive}: \textbf{Awareness}, \textbf{Bias Recognition}, \textbf{Evidence Integration}, \textbf{Reframing}, and \textbf{Agency}. With corresponding interface labels and definitions listed in Table\ref{tab:reflective-categories}.

\begin{table*}[t]
\small
    \centering
    \caption{The five categories of reflective activity derived from cognitive restructuring ~\cite{clark2013cognitive} presented by \systemName.}
    \label{tab:reflective-categories}
    \begin{tabular}{p{0.34\textwidth}p{0.58\textwidth}}
        \toprule
        \textbf{Category / interface label} & \textbf{Conversational activity} \\
        \midrule
        Awareness / Notice the Thought & Naming a salient thought, feeling, or appraisal \\
        Bias Recognition / See a Pattern & Recognizing a recurring or potentially biased thinking pattern \\
        Evidence Integration / Check the Facts & Examining evidence or distinguishing an interpretation from a fact \\
        Reframing / Find a Balanced View & Considering an alternative or more balanced perspective \\
        Agency / Choose My Next Step & Expressing an ownable response or possible next step \\
        \bottomrule
    \end{tabular}
\end{table*}

Each collapsed Progress Scaffold Card displayed the cognitive restructuring category, its plain-language label, and a short description. When the system identified a relevant moment in the user's message, the corresponding card became active and displayed the number of moments assigned to that category. Moment identification was performed by background AI agents. The prompt supplied a definition of each of the five categories, together with a short decision rubric. The model was instructed to judge the participant's latest message and to assign it a category. We iteratively refined the system prompt during the formative pilot study by inspecting whether what was assigned as progress by the system matched participants' perceptions of progress. In addition, the expert reviewer assessed the progress marker categories and what text the system detected and identified as progress, providing validation that the markers accurately captured the types of thought-level progress expected in CR interventions. Expanding the card revealed the conversational text detected as progress alongside an explanation of why the text was detected as progress, under the heading \textbf{Why this may indicate progress}.

\paragraph{Evidence Linking and User Feedback.}
Each expanded card displayed the excerpt from the participant's message on which the interpretation was based. Participants could select \textbf{Highlight in chat} to locate an excerpt in the transcript or select a highlighted passage in the transcript to open its associated card. This bidirectional linking allowed participants to inspect the system's interpretation in the context of their own words~\cite{shneiderman1996eyes}.

Participants could evaluate each detected moment by selecting \textbf{Agree}, \textbf{Not sure}, or \textbf{Disagree} and could optionally provide a note. Feedback was attached to the individual excerpt and stored for analysis, but it did not automatically alter subsequent detections. The controls therefore enabled participants to contest a particular interpretation without suggesting that the model would immediately learn from or resolve that disagreement.

\paragraph{Retrospective Session Summary.}
\label{subsubsec:system-session-summary}
Selecting \textbf{View Summary} opened an AI-generated retrospective account based on the saved transcript and structured session information. In the \systemName condition, the summary first presented a narrative account of the reflection and then a \textbf{What you accomplished} section, which organized supported conversational moments under the five categories.

\subsubsection{Implementation}

The prototype used a TypeScript frontend built with Next.js and React, a Python/FastAPI backend, and a SQLite database. The backend coordinated the conversation and LLM inference through an OpenAI-compatible Responses API client. The system was deployed for experimental purpose using Docker Compose. We provided details on how we prompted the LLM prompts in Appendix ~\ref{appendix:prompt}.

\section{Method}
\label{sec:method}

To examine how access to a progress-sensemaking interface shaped people's perceptions and interpretations of progress toward self-defined psychosocial goals, we conducted a mixed-methods, between-subjects study. Participants were randomly assigned to either a \systemName condition or an active control condition.

\subsection{Participants}

We recruited 30 U.S. adults aged 18 to 70 through Prolific. To be eligible for our study, we required participants to be fluent in reading and writing English and to be able to participate in a Zoom session.
To test our system with its likely users, we recruited participants who were willing to discuss a current psychosocial concern they felt comfortable exploring in a research setting.

The study was approved by an institutional review board (IRB), and all participants provided informed consent. Participants were randomly assigned to one of two system variants ($N=15$ per condition). We report the participant demographics in Table~\ref{tab:participant-characteristics}. Participants were compensated \$30 USD for the approximately one-hour session.

\begin{table}[t]
\centering
\small
\caption{Demographics of our user study participants. Participants with indices starting with X represent users of \systemName, and those with indices starting with Y represent participants in the Active Control condition. We document participants' frequency of AI use for psychosocial support and their perceptions of AI use for general-purpose and psychosocial support (1-5, negative-positive).}
\label{tab:participant-characteristics}
\begin{tabular}{llllll}
\toprule
\multirow{2}[0]{*}{\textbf{P\#}} &
\multirow{2}[0]{*}{\textbf{Gender}} &
\multirow{2}[0]{*}{\textbf{Race/Ethnicity}} &
\multirow{2}[0]{*}{\textbf{AI Use Freq.}} &
\multicolumn{2}{l}{\textbf{AI Attitude}} \\
& & & &
General Use &
Support Use \\
\midrule
X1  & Male   & White
    & Less than once a week
    & \progressplot[6]{5}{1}{5} & \progressplot[6]{4}{1}{5} \\
X2  & Male   & White
    & Less than once a week
    & \progressplot[6]{5}{1}{5} & \progressplot[6]{5}{1}{5} \\
X3  & Male   & Black or African American
    & Less than once a week
    & \progressplot[6]{5}{1}{5} & \progressplot[6]{4}{1}{5} \\
X4  & Female & White
    & More than once a week
    & \progressplot[6]{4}{1}{5} & \progressplot[6]{4}{1}{5} \\
X5  & Male   & White
    & Less than once a week
    & \progressplot[6]{4}{1}{5} & \progressplot[6]{3}{1}{5} \\
X6  & Female & Hispanic or Latino
    & More than once a week
    & \progressplot[6]{5}{1}{5} & \progressplot[6]{5}{1}{5} \\
X7  & Female & White
    & About once a week
    & \progressplot[6]{4}{1}{5} & \progressplot[6]{4}{1}{5} \\
X8  & Female & White
    & More than once a week
    & \progressplot[6]{5}{1}{5} & \progressplot[6]{4}{1}{5} \\
X9  & Female & White
    & More than once a week
    & \progressplot[6]{4}{1}{5} & \progressplot[6]{4}{1}{5} \\
X10 & Female & White
    & More than once a week
    & \progressplot[6]{5}{1}{5} & \progressplot[6]{5}{1}{5} \\
X11 & Female & White
    & About once a week
    & \progressplot[6]{5}{1}{5} & \progressplot[6]{5}{1}{5} \\
X12 & Female & Black or African American
    & Most days
    & \progressplot[6]{5}{1}{5} & \progressplot[6]{5}{1}{5} \\
X13 & Male   & Others
    & About once a week
    & \progressplot[6]{5}{1}{5} & \progressplot[6]{5}{1}{5} \\
X14 & Female & White, Hispanic or Latino
    & About once a week
    & \progressplot[6]{4}{1}{5} & \progressplot[6]{4}{1}{5} \\
X15 & Male   & White, American Indian or Alaskan Native
    & About once a week
    & \progressplot[6]{5}{1}{5} & \progressplot[6]{4}{1}{5} \\
\midrule
Y1  & Female & Black or African American
    & Most days
    & \progressplot[6]{5}{1}{5} & \progressplot[6]{5}{1}{5} \\
Y2  & Male   & Hispanic or Latino
    & Less than once a week
    & \progressplot[6]{4}{1}{5} & \progressplot[6]{2}{1}{5} \\
Y3  & Female & Hispanic or Latino
    & Less than once a week
    & \progressplot[6]{3}{1}{5} & \progressplot[6]{2}{1}{5} \\
Y4  & Male   & White
    & Less than once a week
    & \progressplot[6]{4}{1}{5} & \progressplot[6]{4}{1}{5} \\
Y5  & Male   & White, Asian
    & About once a week
    & \progressplot[6]{5}{1}{5} & \progressplot[6]{5}{1}{5} \\
Y6  & Female & White
    & More than once a week
    & \progressplot[6]{5}{1}{5} & \progressplot[6]{5}{1}{5} \\
Y7  & Male   & Asian
    & About once a week
    & \progressplot[6]{2}{1}{5} & \progressplot[6]{2}{1}{5} \\
Y8  & Male   & White
    & Less than once a week
    & \progressplot[6]{5}{1}{5} & \progressplot[6]{4}{1}{5} \\
Y9  & Male   & White
    & More than once a week
    & \progressplot[6]{4}{1}{5} & \progressplot[6]{5}{1}{5} \\
Y10 & Female & Asian
    & More than once a week
    & \progressplot[6]{3}{1}{5} & \progressplot[6]{4}{1}{5} \\
Y11 & Female & White
    & About once a week
    & \progressplot[6]{5}{1}{5} & \progressplot[6]{4}{1}{5} \\
Y12 & Male   & White
    & Most days
    & \progressplot[6]{2}{1}{5} & \progressplot[6]{2}{1}{5} \\
Y13 & Male   & White
    & Never
    & \progressplot[6]{5}{1}{5} & \progressplot[6]{3}{1}{5} \\
Y14 & Female & Black/African American, White
    & About once a week
    & \progressplot[6]{5}{1}{5} & \progressplot[6]{5}{1}{5} \\
Y15 & Female & White
    & More than once a week
    & \progressplot[6]{5}{1}{5} & \progressplot[6]{5}{1}{5} \\
\bottomrule
\end{tabular}
\end{table}

\subsection{Conditions}

\begin{figure}
    \centering
    \includegraphics[width=.83\linewidth]{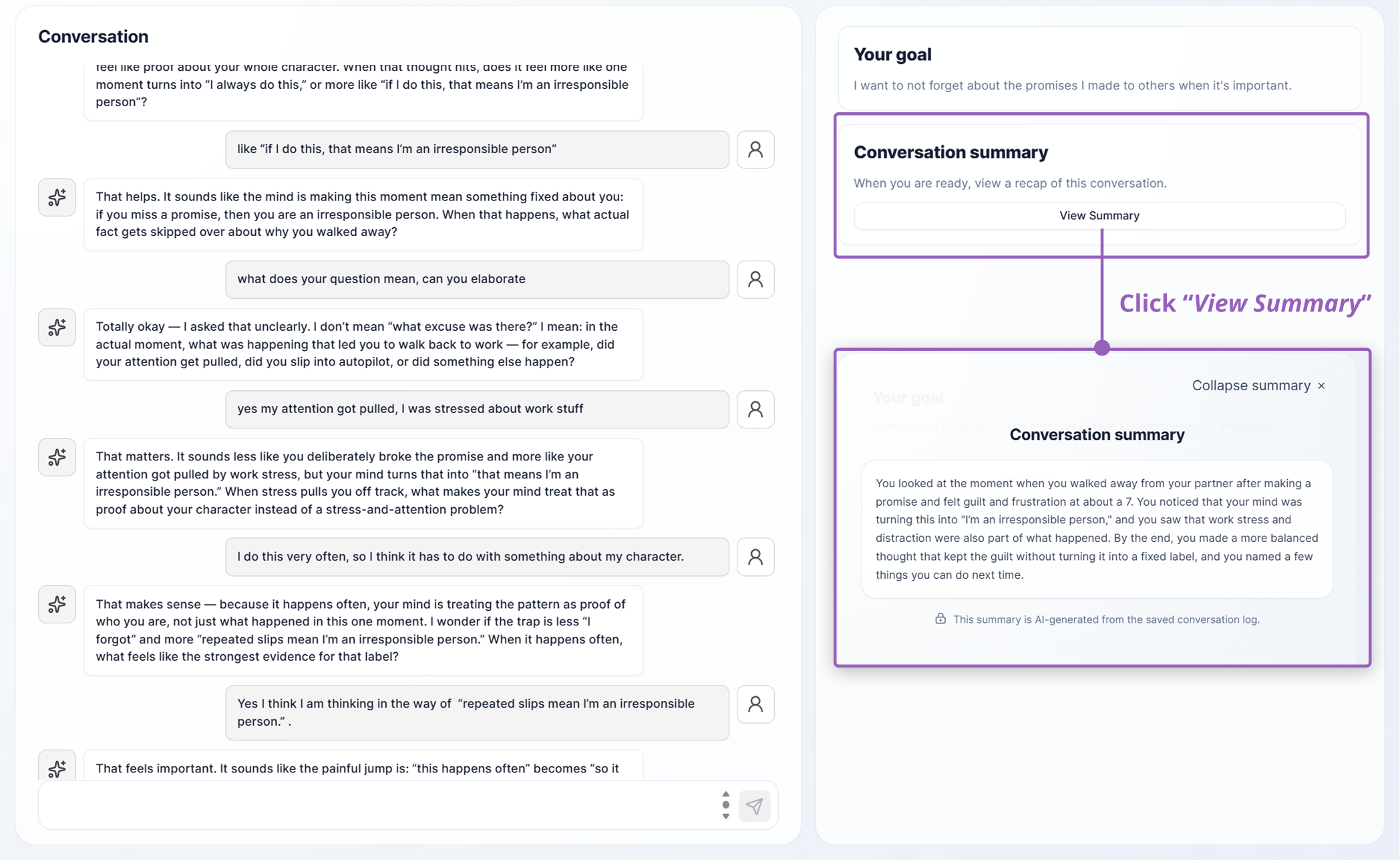}
    \caption{Interface for the \baseline condition. We removed \systemName's additional feature related to in-the-moment progress visualization. Upon clicking on ``view summary,'' the \baseline condition displays a plain summary without the reflective categories.}
    \label{fig:baseline}
    \Description{}
\end{figure}

To understand how the progress-sensemaking scaffold of \systemName influenced understanding of psychosocial goal progress, we compared \systemName with an \textbf{active control condition}. We assigned 15 participants to each of the two system variants.

In the \textbf{active control condition}, the right-hand panel displayed only the participant's goal (see Fig.~\ref{fig:baseline}). 
Compared with \systemName, participants did not see Progress Scaffold Card, category counts, highlighted excerpts, or interpretation-feedback controls during the conversation. After the conversation, participants were instructed to click ``View Summary'' to see a narrative recap similar to \systemName. 
In \systemName, the summaries were organized by the progress marker categories; in \baseline, they were not. 
To sum up, the \baseline condition used the same cognitive restructuring prompts and included a similar retrospective summary as \systemName, but without the in-the-moment progress markers and summary categorizations, thereby allowing us to focus on understanding the effect of the progress-sensemaking scaffold.

We selected an active control over a no-support control (i.e., no conversational AI support) as the baseline for three primary reasons. First, we are specifically able to estimate the effects of \systemName as a novel progress sensemaking system; a no-support control would not isolate these effects, and instead would estimate the effect of \textit{both} progress sensemaking and conversational AI mental health support. Second, the active control condition reflects a common conversational AI pattern that combines an interactive dialogue with basic retrospective support, such as a conversation summary~\citet{kim2024mindfuldiary}. Third, the \baseline condition retained the same goal-formation flow, prototype walkthrough, LLM-guided conversation, and narrative recap as the \systemName condition, allowing us to more directly examine what the additional \systemName scaffold contributed to users' perceptions and interpretations of progress.

\subsection{Procedures and Tasks}
\label{subsec:procedure-task-study}

Each approximately one-hour study session was conducted remotely via Zoom. 

\paragraph{Introduction and pre-prototype questions.}
After providing informed consent and completing the background questionnaire, participants were introduced to the concept of psychosocial goals. The interviewer asked how they normally recognized progress toward such goals, what kinds of changes they looked for, and what made psychosocial progress difficult to notice or track.

\paragraph{Goal formation and reflection task.}
The researcher first introduced the study task, after which participants completed the three-step goal-formation flow with procedural guidance from the researcher. Participants then received a brief walkthrough tailored to the features available in their assigned condition. Before beginning the conversation, participants rated their current mood, anxiety, and stress using three single-item, five-point scales. They then engaged in an LLM-guided cognitive restructuring conversation about their self-defined goal. Participants were instructed to think aloud throughout the interaction by verbalizing what they noticed, expected, or found confusing as they used the system. At the end of the conversation, they rated their mood, anxiety, and stress again using the same items. Participants then opened and reviewed the retrospective summary available in their assigned condition.

\paragraph{Post-use survey and interview.}
Participants next completed the post-use questionnaire in Qualtrics. The interviewer then conducted a semi-structured interview building on participants' experiences with the system and their think-aloud comments.

\paragraph{Live supervision and safety.} Under our IRB-approved protocol, every session was supervised live by a trained interviewer who observed the participant's conversation. The interviewer was responsible for protocol adherence and participant safety, and was authorized to pause or discontinue a session if a participant showed signs of distress or attempted to use the tool outside its intended scope. Responsibility for handling risk during the study therefore rested with the researcher rather than with the system: the policy described in \S\ref{subsubsec:system-conversation-safety} governed the chatbot's own responses, but was not relied upon as the study's safeguard. In practice, the researcher did not observe any signs of distress/out-of-scope use, no session was paused or discontinued, and the model correctly did not identify acute risk in any session.

During consent, participants were told that they were interacting with an AI tool rather than a person, that the tool was for research purposes and not a substitute for professional support, and that their inputs and the system's responses would be processed by a third-party model provider. Individuals who were not comfortable with these terms could decline to participate.

\subsection{Data Collection}

During the interviews, we collected from participants self-reports to measure differences between \systemName and the active control, and qualitative data from semi-structured interviews.

\paragraph{Self-Reports.}
Before and after using the system, participants completed a brief EMA-style assessment~\cite{shiffman2008ecological} of their current affective state. They rated their mood from 1 (Worst) to 5 (Best), anxiety from 1 (Low anxiety) to 5 (High anxiety), and stress from 1 (Low stress) to 5 (High stress). Twenty-nine participants completed both assessments; one participant did not provide the post-system ratings.

Participants completed additional self-reports after using the prototype. All items used five-point Likert-type agreement response options ranging from 1 (Strongly disagree) to 5 (Strongly agree). We first assessed usability with the 10-item System Usability Scale (SUS)~\cite{brooke1996sus}. To assess the system's working alliance, we included seven items from the Working Alliance Inventory--Short Revised (WAI-SR)~\cite{hatcher2006development}: four goal-subscale items and three of the four task-subscale items. One task-subscale item was omitted due to a technical error. Following a similar approach to prior work assessing users' alliance with conversational AI mental health support tools~\cite{darcy2021evidence}, items were assessed by replacing the word ``therapist'' with ``system.'' We focus on these items, while omitting administering the WAI-SR bond subscale, as our research questions focus on how a system can support users in understanding progress towards attaining their goals, and we did not intend for participants to form a bond with the system~\cite{bordin1979generalizability}.

We also included six relevant items from the Technology-Supported Reflection Inventory (TSRI)~\cite{bentvelzen2021development} to assess the extent to which the system supported insight, exploration, consideration of change, and reflection on participants' goals. The six items were adapted from the original TSRI. We omitted the comparison subscale, which concerns reflecting on personal data with others, discussing it with others, and relating it to others' data~\cite{bentvelzen2021development}. Neither condition provided features for sharing or comparing personal data with other people; instead, both supported individual reflection on participants' own concerns and conversations.

Finally, we asked six additional questions motivated by prior measures to examine perceived goal attainability, understanding of what progress toward a goal could look like, motivation to continue working toward the goal, perceived competence to attain the goal, perceived pathways toward the goal, and current goal attainment. The questions were motivated by prior measures of goal attainability~\cite{schimpf2026supporting}, the perceived-impact items in Section F of the Mobile App Rating Scale~\cite{stoyanov2015mobile}, the perceived-competence component of the Intrinsic Motivation Inventory~\cite{mcauley1991self}, and the agency component of the State Hope Scale~\cite{snyder1996development}.

\paragraph{Semi-structured interviews.}
Following the post-use questionnaire, we conducted semi-structured interviews to understand how participants interpreted progress toward their self-defined psychosocial goals and how the system shaped that sensemaking process. Questions addressed how participants recognized progress, whether particular conversational moments felt meaningful, how well the system's representations fit their experiences, preferences regarding feedback timing, emotional responses, cognitive effort, potential real-world use, and desired changes. Interviewers asked follow-up questions based on participants' responses.

\subsection{Quantitative Analysis}
\label{subsec:quantitative-analysis}

To support analysis of the effect of \systemName, we used Bayesian regression models, implemented with the Python package \texttt{Bambi} (v0.16.0) and reported with package \texttt{ArViz} (v0.23.4). Instead of Frequentist approaches, which make inferences from collected data through hypothesis tests and confidence interval reporting~\cite{gelman2017beyond}, we chose Bayesian modeling because it provides richer interpretative information and is better suited for smaller sample sizes~\cite{van2014gentle}. Specifically, Bayesian modeling estimates the full posterior distributions of responses, allowing us to make probabilistic statements about both the direction and plausible magnitude of an effect. For example, we can report the posterior probability that an effect is positive and is likely to exceed a meaningful threshold. Such properties 

better supports our paper's goal of understanding the possible effects of \systemName's design.

Moreover, because we hypothesized that participants' familiarity and attitudes toward AI may influence their perceptions of \systemName condition, we modeled their self-reported frequency of AI use (\textit{AI\_freq}), perceptions of AI (\textit{AI\_perc}), and attitudes toward AI for psychosocial support (\textit{AI\_support}), which were all measured during the screening survey. Accordingly, we treat those variables as covariates in our Bayesian models. We treated the covariates as continuous and centered the responses around 0 (denoted with \textit{\_c}). We modeled those responses as continuous rather than ordinal to maintain model parsimony given the limited sample size.

We chose different categories of Bayesian regression models based on attributes of the outcome variables. All outcome variables were analyzed as separate items because they were selected to capture conceptually distinct aspects of participants' experiences rather than to jointly measure a single underlying construct. For all the measures that used a 5-point Likert Scale, we used Bayesian Ordinal Regression to account for the ordered, discrete nature of Likert responses. Specifically, given an ordinal outcome variable \textit{measure}, we create the following model:
\begin{equation}
\text{<\textit{measure}>} \sim 0 + \text{\textit{condition}} + \text{\textit{AI\_freq\_c}} + \text{\textit{AI\_perc\_c}} + \text{\textit{AI\_support\_c}}
\end{equation}
where condition is either the \baseline (referent, coded as 0) or \systemName condition (coded as 1). The equation was written in the syntax of the Python package \texttt{Bambi}, where the term 0 indicates the ``no intercept'' setup as indicated by \texttt{Bambi}'s documentation\footnote{\url{https://bambinos.github.io/bambi/notebooks/ordinal\_regression.html}}. For continuous outcome measures (i.e., SUS score), we used a similar model (with intercept, i.e., without the 0) with Bayesian Linear Regression.

Additionally, to capture the pre-post difference in mood, stress, and anxiety, we used a Difference-in-Differences model. We introduced a \texttt{time} variable and coded pre-test as 0, and post-test as 1. We used the following model:
\begin{equation}
\text{<\textit{mood/stress/anxiety}>} \sim 0 + \text{\textit{condition}} \times \text{\textit{time}} + \text{\textit{AI\_freq\_c}} + \text{\textit{AI\_perc\_c}} + \text{\textit{AI\_support\_c}}
\end{equation}
where we focus on interpreting the interaction term $\text{\textit{condition}} \times \text{\textit{time}}$. Since mood, stress, and anxiety were measured on a 5-point Likert scale, we used Bayesian Ordinal Regression for these outcome variables. We used automatically selected \textit{weakly informative priors} from the package \texttt{Bambi} for all the models above to regulate our models against extreme estimates, without forcing the data toward any particular outcome.

When reporting the effects, we present $\beta$ and 95\% Equal-tail Credible Intervals (CrI, the Bayesian equivalent concept of Confidence Interval). Specifically, $\beta$ is the posterior mean of the regression coefficient (on the log-odds scale for Ordinal Regression), with a positive value indicating a positive association between the predictor and the outcome. The \baseline condition is used as the referent when reporting $\beta$.  CrIs excluding zero indicate that the central 95\% of posterior mass lies entirely on one side of zero (similar to the Frequentist concept of ``significance''), while intervals spanning zero indicate greater uncertainty about the direction of the effect. Unlike Frequentist confidence intervals, which are interpreted based on how often similarly constructed intervals would contain the true effect across repeated studies, CrIs can be interpreted directly as there is a 95\% probability that the effect lies within the reported range, given the model, priors, and observed data. To ensure the statistical reporting is easier to interpret, we choose to report in a style similar to Frequentist regression analysis. However, 
 
we advise readers to interpret the results by considering both the magnitude of the effect and the uncertainty represented by the credible interval.

\subsection{Qualitative Analysis}
\label{subsec:qualitative-analysis}

Interview and think-aloud transcripts were checked for accuracy and de-identified by the first author before analysis. The first two authors analyzed the transcripts using a thematic analysis approach informed by~\citet{braun2006thematic}. The authors first familiarized themselves with the data by reading the transcripts and writing initial analytic notes. We then conducted open coding in a shared spreadsheet, guided by the research questions and literature review. We then consolidated the code excerpts into a shared spreadsheet that contained the coded excerpts, code labels, and code definitions for the document. The authors met regularly to compare their interpretations, discuss differences, and refine the codes and their definitions using the spreadsheet. We then grouped related codes into themes and examined the relationships among them. Through iterative discussion, the authors refined and named the final themes and selected representative excerpts to illustrate each theme. We report the findings through this analysis in Section~\ref{subsec:findings-pyschosocial-progress-scaffold}. An overview of the themes, subthemes, and illustrative quotes is provided in Appendix~\ref{appendix:themes}.

\section{Findings}
\label{sec:findings}

In this section, we first report the quantitative findings in \S\ref{subsec:findings-quant} on measures of alliance, reflection, attainability, change in affect, and usability. We then turn to the qualitative findings. We describe how the progress scaffold helped participants notice, make sense of, and trace moments of progress in their own words (\S\ref{subsec:findings-pyschosocial-progress-scaffold}); how they distinguished progress within the conversation from progress toward their broader goal (\S\ref{subsubsec:findings-movement-versus-change}); and how they weighed real-time feedback against a retrospective summary (\S\ref{subsec:findings-timing}).

\begin{figure}
    \centering
    \includegraphics[width=1\linewidth]{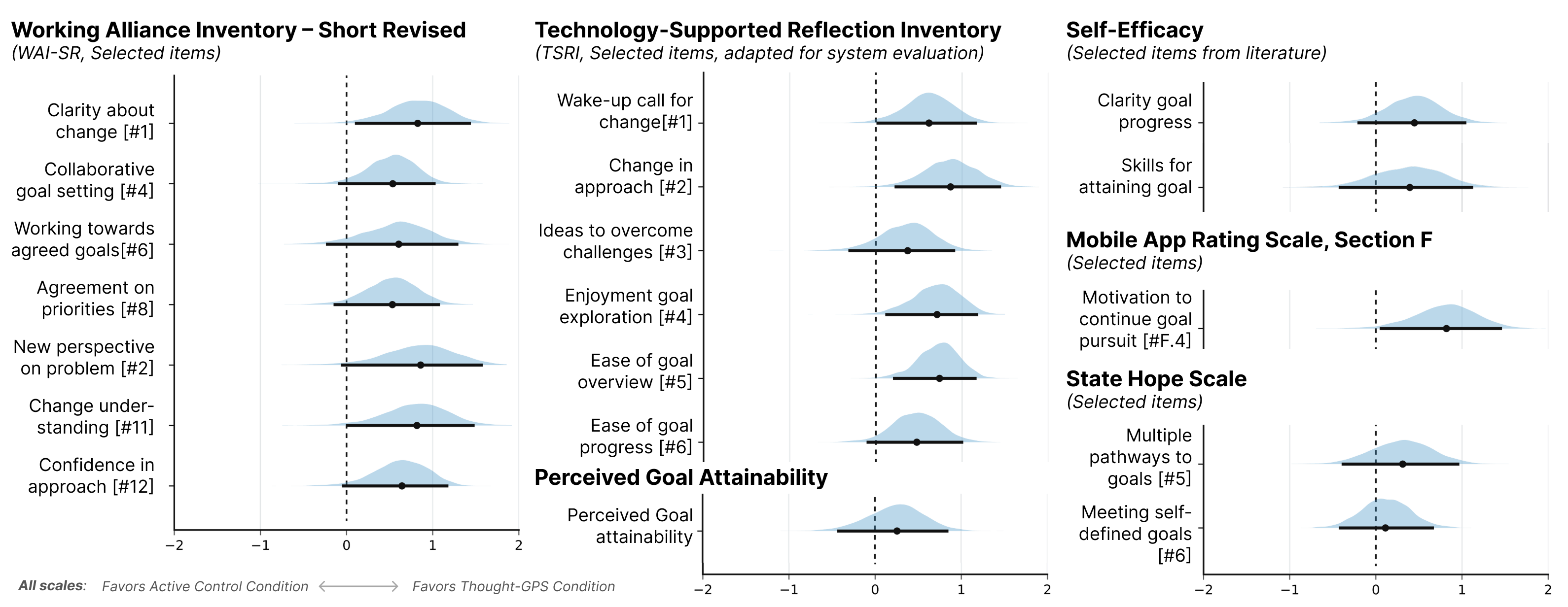}
    \caption{Posterior densities of the contrast in expected Likert ratings between the \systemName and the \baseline, on the original five-point response scale. Positive values indicate favoring \systemName over \baseline. The dot indicates the posterior median of the difference distribution, while the horizontal line shows its 95\% credible interval. The numbers in the square brackets correspond to the question numbers in the original scale. We document the original questions in Appendix~\ref {appendix:survey-questions}.}
    \label{fig:quantresult}
    \Description{}
\end{figure}

\begin{figure}
    \centering
    \includegraphics[width=0.25\linewidth]{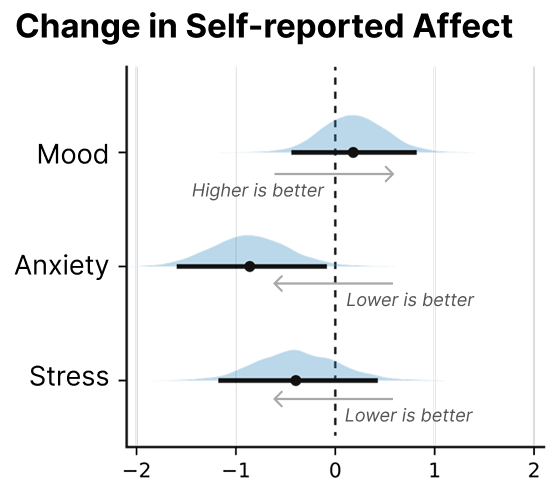}
    \caption{Posterior densities of the difference in difference in expected Likert ratings: \((\text{\systemName}_{\mathrm{post}}-\text{\systemName}_{\mathrm{pre}})-(\text{\baseline}_{\mathrm{post}}-\text{\baseline}_{\mathrm{pre}})\), transformed to the original five-point scale for mood, anxiety, and stress. Dots indicate posterior medians, and horizontal lines show 95\% credible intervals.}
    \label{fig:quant-affect}
    \Description{}
\end{figure}

\subsection{Quantitative Findings}
\label{subsec:findings-quant}

Across several measures of alliance, reflective engagement, and goal motivation, we found preliminary evidence of more favorable ratings for \systemName than for the \baseline condition (see Fig.~\ref{fig:quantresult}). 

Participants using \systemName were more likely to report clarity about how they might change from using the system (\bstat{1.64}{0.15, 3.29}, WAI-SR Item 1). 
In addition to clarity, participants felt more engaged reflecting with \systemName over \baseline, including a sense that the system has led to a realization to make changes (\bstat{1.49}{0.01, 3.02}, TSRI Item 1) and prompted a reflection on how they approach things (\bstat{1.78}{0.38, 3.34}, TSRI Item 2). 
Participants using \systemName also reported greater enjoyment in exploring their goals (\bstat{1.71}{0.23, 3.27}, TSRI Item 4) and greater ease in obtaining an overview of those goals (\bstat{2.02}{0.46, 3.73}, TSRI Item 5). 
Participants in the \systemName condition also reported stronger motivation to continue working toward their goals with \systemName than in the \baseline condition (\bstat{1.51}{0.07, 3.02}, MARS Section F Item 4).

\systemName also showed potential for supporting participants' affect regulation (see Fig.~\ref{fig:quant-affect}). Participants using \systemName reported a larger reduction in anxiety over the course of the session (\bstat{-2.13}{-4.16, -0.22}). While the effects on mood and stress are not considered meaningfully different, both trend in a direction favoring the \systemName condition, with improved mood (\bstat{0.58}{-1.32, 2.66}, with a 71.9\% chance that \systemName led to greater improvement) 
and reduced stress (\bstat{-0.92}{-2.78, 0.98}, with a 83.2\% chance that \systemName led to greater reduction). Although the posterior estimates generally favored \systemName, the credible intervals were wide, indicating uncertainty about the magnitude of these differences. 

Additionally, despite \systemName having more features than \baseline, both conditions recorded relatively high usability scores ($Mean\_SUS_{\systemName}$=86.17, $Mean\_SUS_{\baseline}$=83.33, \bstat{1.32}{-9.42, 11.93}). Taken together, these item-level estimates indicate preliminary evidence that adding in-the-moment conversation progress visualizations alongside chatbot-mediated psychosocial support may improve clarity about change and engage participants in reflection.

\label{subsec:findings-qua}
We also found qualitative evidence that further contextualizes these quantitative findings, which we describe in the remaining findings sections. First (\S\ref{subsec:findings-pyschosocial-progress-scaffold}), participants described the scaffold as making subtle reflective moments visible, supporting deeper sensemaking, and allowing the system's interpretations to be traceable towards their conversation data. Second (\S\ref{subsec:findings-conversation-progress-VS-goal-progress}), conversational progress was distinguished from progress toward broader psychosocial goals. Participants often look for longitudinal, behavioral, or real-world evidence before concluding that the goal itself has advanced. Third (\S\ref{subsec:findings-timing}), participants described feedback timing as a trade-off: in-the-moment cues made emerging changes salient but could interrupt the conversation, whereas retrospective feedback supported synthesis across the interaction. 

\subsection{Using psychosocial progress scaffold for understanding, evaluating, and revisiting progress}
\label{subsec:findings-pyschosocial-progress-scaffold}

Participants described three related ways in which the scaffold supported their understanding of progress towards their goal: by making reflective moments visible (\S\ref{subsubsec:findings-reflection}), supporting deeper interpretations about those moments (\S\ref{subsubsec:findings-sensemaking}), and by providing transparency of the conversation excerpt sources (\S\ref{subsubsec:findings-transparency}). 

\subsubsection{Reflection: Progress scaffolds turned previously unnoticed moments of progress into visible evidence}
\label{subsubsec:findings-reflection}

Participants often described psychosocial progress as difficult to recognize. They characterized it as subjective and easily obscured by memory or short-term emotional fluctuations. Despite these challenges, participants using \systemName turned previously hard-to-notice moments of progress into visible evidence. For example, X2 explained:
\begin{quote}
    \textit{It just made it a lot more visible as it was happening, versus if I was just kind of thinking about it myself, it wouldn't have been so apparent to me.} - X2
\end{quote}
X1 shared a similar sentiment, viewing the category-organized Progress Scaffold Cards, stating \textit{``I normally wouldn't catch stuff like that (category-organized Progress Scaffold Cards).''}

The progress scaffold, particularly the linked highlights and Progress Scaffold Cards, made otherwise vague signs of progress visible by organizing them into progress categories and connecting each interpretation to a specific statement. Participants described these representations as helping them separate important reflective moments from the surrounding conversation, functioning like a set of highlighted notes that they could quickly locate and revisit. For example, X4 said that a card \textit{``breaks it down into the evidence, and then, like, a step, and it also shows the progress, so it's very, like, evidence-based.''} The scaffold therefore did more than label a conversational moment as progress. It made changes in participants' thinking perceptible while the conversation was unfolding.

Making progress visible may also have supported perceived engagement. X2 said that the visible feedback, including highlights and Progress Scaffold Cards that lit up, made them want to \textit{``engage more and provide more information.''} X5 described each new Progress Scaffold Card as \textit{``a little dopamine rush,''} and X6 said that the real-time feedback \textit{``gets my wheels turning.''} Those results also corroborated the quantitative finding, with items related to the enjoyment of exploring (TSRI Item 4) with the systems and motivation to continue exploration with the system (MARS Section F Item 4) both displayed credibly different result comparing \systemName with the \baseline.

These accounts suggest that seeing progress emerge during the conversation could make the interaction feel more engaging and motivate participants to continue reflecting.

\subsubsection{Sensemaking: Conversing with the chatbot supported deeper sensemaking}
\label{subsubsec:findings-sensemaking}

Making reflective moments visible was the scaffold's most immediate contribution, but visibility alone did not help explain how participants understood those moments. We next examine how the conversation and the additional scaffold supported participants in interpreting and elaborating on possible progress.
Across both conditions, participants described the identification of particular moments as increasing their awareness of progress. In the \baseline condition, for example, Y1 said that the summary allowed them to see \textit{``what I accomplished during the conversation''} and take away the information easily. Y7 similarly explained that the conversation made progress clearer because \textit{``I've never really said some of these things out loud before.''} These accounts indicate that the shared reflective process itself supported progress awareness, even without in-the-moment progress cues.

Participants in the \systemName condition reported stronger post-use clarity and understanding impact. On the survey item asking whether the system increased their understanding of what progress toward the goal could look like, participants in the \systemName condition more often agreed or strongly agreed than participants in the \baseline condition. Specifically, we observed a credible difference in how clearly participants felt they understood in how they could make changes (WAI-SR Item 1) and how they might approach situations differently (TSRI Item 2) when interacting with \systemName compared with \baseline.

The qualitative data suggested why these differences might have occurred. The visible cues may have supported deeper sensemaking. The cards and highlights prompted some participants to pause, reconsider an interpretation, and elaborate on what they had said. X3 attributed this process to the side panel: highlighting worries that were not supported as facts \textit{``pushes me to look at it just like that. Like, what I'm contemplating in my head over and over isn't a fact, so what can I do to address it?''} X4 said that the Progress Scaffold Cards \textit{``had me go deeper into the emotions,''} while X6 said that the cards revealed that their statements were \textit{``so much deeper''} than they had initially appeared. These accounts suggest that the scaffold supported not only noticing an insight, but also inspecting, reflecting on, and refining it.

Thus, participants in both conditions reported benefits from reflection, while those with access to the additional scaffold reported greater clarity on how they might change after using the system. 

\subsubsection{Transparency: Source-Linking as Supports for connecting back to the conversation}
\label{subsubsec:findings-transparency}

The ability to evaluate the system's interpretations became increasingly important as participants interacted and engaged with progress scaffold moments. Source-linking supported this evaluation by connecting each interpretation to the participant's own conversational evidence.

The system's interpretation was most convincing when participants could trace back the grounding in their own words. X3 stated \textit{``Once I read the card, I would go back to the conversation to kind of remind myself of what I said.''} This link allowed participants to treat the system's interpretation as a proposition they could inspect rather than a conclusion. X4 similarly described the cards as breaking an interpretation down into its evidence, which made the representation feel evidence-based to them. The ability to revisit the original statement supported both understanding and evaluation of the system's interpretation.

Participants did not report disagreeing with the markers. When asked whether the system had marked something as progress that they disagreed with, participants expressed their agreement with the markers. X1 stated, \textit{``I didn't disagree with any of it. It all was legit to me. Well explained and everything.''} X5 similarly explained, \textit{``The things it was calling out and highlighting were things it should have been highlighting, and that helped me trust it more.''} Others, when asked whether the markers were accurate or whether the system had missed anything, described them as well-targeted rather than as mistaken. Few recorded these judgments through the interface's optional Agree, Not sure, or Disagree ratings, suggesting that participants either found the markers agreeable or were not focused on using these judgment features.

\subsection{Progress Within the Conversation versus Progress Toward the Goal}
\label{subsec:findings-conversation-progress-VS-goal-progress}

While visibility, deeper sensemaking, and source-linking made changes within the conversation more available for interpretation and evaluation, participants did not equate such clarity with progress towards their broader psychosocial goals (\S\ref{subsubsec:findings-movement-versus-change}). Instead, participants perceived the newly found understanding as starting points for longitudinal tracking (\S\ref{subsubsec:findings-longitudinal}).

\subsubsection{A single session offered a snapshot, prompting requests for longitudinal tracking}
\label{subsubsec:findings-longitudinal}
Some participants felt that a single session could provide a useful snapshot of progress, but longitudinal tracking was preferred for showing change persisted over time. Participants understood progress toward their psychosocial goals as something that unfolds over weeks and months, well beyond the horizon of a single session. Several participants across both conditions requested functionalities such as chat memory, chat history, or longitudinal tracking across conversations. X11 characterized the single-session interaction as \textit{``one frame of a 67-year-old movie of my life,''} emphasizing that the system had only a snapshot of the relevant context. X2 wanted confidence that they could \textit{``go back to this in the future''} and review stored history to track progress. X13 imagined comparing summaries from different days and weeks and having the system compress them \textit{``so that you're moving towards something rather than just ... an extract of a conversation.''}

These requests clarify a limit for one-session progress feedback. The scaffold could visualize evidence of progress in the current interaction, but it could not determine whether participants later thought differently, whether an emotional shift persisted, or whether the external situation changed. Longitudinal views were therefore requested because participants wanted to see the recorded progress as a reminder, as well as how the conversational progress evolves over time.

\subsubsection{Participants distinguished movement within the conversation from a change in the underlying situation}
\label{subsubsec:findings-movement-versus-change}

The desire for longitudinal evidence reflected a broader distinction in how participants evaluated progress: movement within a conversation could be meaningful without being sufficient to demonstrate advancement toward the underlying goal.
Although participants recognized the moments identified by the scaffold as progress within the conversation and generally agreed with the system's interpretations, they did not always view these changes as evidence that their broader goal had advanced. X5 captured this difference directly:

\begin{quote}
    \textit{I don't feel like I progressed to a goal yet... but I feel like there's a plan.} - X5
\end{quote}

Participants' own criteria for judging what counts as progress further reflected this difference. Whereas some participants treated emotional relief as a meaningful signal, others looked for behavioral change or real-world outcomes, such as forming concrete plans, completing steps, observing change over time, or receiving feedback from other people. Y10 drew a clear boundary between clarifying a goal through conversation and making progress outside it: \textit{``The actual ... progress ... is me doing it outside of the chatbot interaction. My goal was to ... meet up with my family more during the weekends. I can't really do that with the chatbot.''} Changes in awareness or interpretation could therefore be valuable without necessarily leading participants to conclude that broader progress had occurred.

The \systemName interpreted progress aligned most directly with participants who regarded changes in thinking---such as understanding their reactions, reconsidering assumptions, or working through rumination---as meaningful progress toward their goals. For these participants, changes in thinking could themselves constitute meaningful progress toward their goals.  For example, X7 described that progress did not necessarily require the complete disappearance of anxiety, but could instead involve greater agency over how they responded to it:

\begin{quote}
    \textit{Progress doesn't mean that ... [the] cloud of anxiety isn't there anymore. It's more [that] you have the autonomy to kind of change that thinking, even when it does show up.} - X7
\end{quote}

For goals that depended on actions or changes outside the conversation, participants wanted support beyond what the scaffold represented. When the system treated a real financial threat primarily as a thinking pattern, X12 objected:

\begin{quote}
    \textit{I don't have any more money, and now I'm moved to borrowing money from friends ... that's a worst-case scenario.} - X12
\end{quote}

X11 similarly said that the system helped them state positive possibilities but \textit{``didn't give me the advice on how to do this.''}
Together, these accounts show that participants could value changes that occurred within the conversation without treating them as evidence that their broader goal had advanced. The scaffold made thought-level progress visible, while some participants relied on subsequent plans, actions, and real-world outcomes when judging broader progress. We return to this distinction in the Discussion.

\subsection{When Progress Scaffolds Should Appear: Salience, Synthesis, and the Cost of Switching}
\label{subsec:findings-timing}

In-the-moment markers and the retrospective summary were described as doing different work. Participants said the live markers made emerging changes salient while the conversational context was still available, whereas the summary helped them synthesize the interaction as a whole (\S\ref{subsubsec:findings-timing-salience}). Preferences between the two were divided (\S\ref{subsubsec:findings-timing-conversation-continuity}), with several participants pointing to the attentional cost of turning away from the conversation to inspect live feedback (\S\ref{subsubsec:findings-context-switching}).

\subsubsection{In-the-moment feedback provided salience, whereas retrospective feedback provided synthesis}
\label{subsubsec:findings-timing-salience}

Participants described the two feedback timings as supporting distinct but complementary forms for understanding progress. In-the-moment highlights located a statement in its conversational context and surfaced a possible progress interpretation while the exchange was still underway. X6 explained:

\begin{quote}
    \textit{It made me choose my words more decisively... narrow down and kind of get to the point.} — X6
\end{quote}

X9 expressed a similar idea, explaining how real-time progress insights could facilitate the ongoing conversation thought progress:

\begin{quote}
    \textit{I think it's good in real time. Because... that's when you have to attack the thoughts... as you're having them, rather than after the fact.} — X9
\end{quote}

Compared with progress visualized in real time, retrospective summaries helped participants combine scattered moments into a coherent account. X7 valued both forms: they \textit{``enjoyed the real-time''} feedback but also \textit{``really enjoyed the overall summary ... to kind of regroup it all together.''} X13 similarly preferred an overall summary to reviewing individual evidence cards because the summary made the conversation feel coherent.
These findings imply that the two forms of feedback are not interchangeable. Immediate feedback answers, ``What just happened, and why might it matter?'' whereas retrospective feedback answers the question ``What did this conversation add up to?''

\subsubsection{Feedback timing created a trade-off between progress awareness and conversational continuity}
\label{subsubsec:findings-timing-conversation-continuity}

Participants' timing preferences reflected a trade-off between making progress salient and preserving continuity of the conversation. Participants did not converge on one ideal timing for displaying the progress scaffolds. Among participants in the  \systemName condition, preferences were roughly evenly divided between preferring in-the-moment feedback versus reviewing feedback after the conversation. The formative sessions showed similar variation. P1 expected to review feedback more often after the conversation, but also found the live cues helpful. P3 preferred lightweight highlights during the conversation and more detailed review afterward. Those who preferred real-time feedback valued knowing that progress was occurring while the relevant context was still available.

Others preferred to complete the CBT-style conversation before reflecting on progress. X3 said, \textit{``I would probably prefer it at the end ... because I could see myself getting distracted.''} Y14, who had only seen the retrospective summary and was asked afterward to imagine a live version, anticipated that it would cause them to \textit{``focus more on the feedback and not the conversation.''} The split therefore reflected more than a preference about timing: it depended on whether participants wanted to stay immersed in the conversation or step back to evaluate it. These accounts suggest that users should be able to decide when to inspect progress feedback based on their reflective needs and the context of the conversation.

\subsubsection{The additional scaffold did not change the overall usability, but attending to it introduced context switching for some participants}
\label{subsubsec:findings-context-switching}

The attentional costs associated with live feedback were not reflected in participants' overall assessments of system usability. Both \systemName and \baseline received general compliments in their ease of use, both quantitatively (both systems scored an average of over 80, equivalent to A grade usability or above) and qualitatively (i.e., describing the system using words such as ``easy to use'', ``clear'', or ``user-friendly'').

High usability, however, did not mean that attending to the live scaffold came without effort. Several \systemName participants described switching their attention between continuing the conversation and inspecting new highlights or Progress Scaffold Cards. X3 said that this switching \textit{``caused me to have to divert my attention from the conversation''} and made parts of the interaction feel \textit{``a little bit more mechanical.''} X13 described the difficulty of \textit{``jumping back between sitting outside of it versus being in it''} during an emotional conversation. X15 similarly explained that having to open every highlight \textit{``takes away from keeping the conversation flow.''} These accounts characterize the switching between conversing and inspecting feedback as a cost of real-time progress scaffolds.

\section{Discussion}
\label{sec:discussion}

Our findings suggest that \systemName helped participants recognize and make sense of possible progress during reflective conversations. Compared with \baseline, participants reported greater clarity about how they might change, stronger reflective engagement and goal motivation. Qualitative accounts showed how linked highlights and Progress Scaffold Cards made thought-progress visible and allowed participants to inspect interpretations grounded in their own words. Participants nevertheless distinguished these shifts from broader goal progress and wanted to revisit them over time. They also described complementary benefits and tensions in feedback timing: live markers supported immediate awareness but could interrupt conversational flow, while retrospective summaries helped bring the interaction together.

This discussion focuses on three questions: what kind of progress a conversational system can credibly claim to support, how that progress should be surfaced without displacing the user's own interpretation or interrupting the reflection it is meant to aid, and how to make the interpretation traceable.  

\subsection{Distinguishing Thought-Level Progress from Behavior-Level Progress}
\label{subsec:discussion-thought-vs-behavior}

In this section, we distinguish thought-level progress from behavior-level progress and discuss how this distinction defines the scope of conversational scaffolds and their potential applications.
Our findings in Sections ~\ref{subsubsec:findings-reflection} and ~\ref{subsubsec:findings-movement-versus-change} highlight differences in how participants interpreted the scaffolds as progress towards their specific psychosocial goals. 
Many participants were more accustomed to viewing progress as behavioral changes they would need to make to attain their goals~\cite{ekhtiar2023goals}. As introduced in \S\ref{sec:introduction}, we distinguish thought-level progress — shifts in how people notice, evaluate, and reframe their thoughts — from behavior-level progress, which occurs downstream and has been the focus of existing goal tracking systems in HCI.

This distinction can be interpreted similarly to proximal and distal outcomes: proximal outcomes are the direct intervention targets (e.g., completing a mood check-in), whereas distal outcomes are downstream and are often what interventions intend to change (e.g., sustained reduction in depressive symptoms) \cite{nahumshani2015building}.
Similarly, in psychotherapy, in-session change (e.g., reframing a thought) has been linked to, but distinguished from, overall treatment outcomes (e.g., reducing anxiety symptoms) \cite{muran1995linking}.
In Section~\ref{subsubsec:findings-movement-versus-change}, participants required different kinds of evidence before recognizing progress: some emphasized emotional or physical relief, while others wanted to take concrete actions, observe specific outcomes, or receive feedback from other people. An accurate indication of thought-level progress, therefore, did not always translate into a perceived sense of progress towards goals.

We argue that the thought-level/behavior-level distinction provides a useful boundary for the design of future systems that expand the affordances of \systemName. Novel systems that extend \systemName into the longitudinal context could first use thought-level progress scaffolds to give individuals concrete evidence of how reframing thoughts can support psychosocial goal attainment~\cite{strunk2007patients, gomezpenedo2020theory}. 
These systems could then support downstream behavior-level progress tracking---similar to Bloom ~\cite{jorke2026bloom} or GoalTrack~\cite{lee2024goaltrack}---as individuals take actions to attain their goals.

This framing positions the progress markers presented in \systemName as a first step towards attaining psychosocial goals.

The distinction also clarifies the role of the CBT-style cognitive restructuring conversation used in our prototype. Cognitive restructuring may be best suited to goals whose difficulty lies in how a person interprets a situation, such as rumination or anxious thinking. For such goals, noticing a thought pattern or arriving at a more balanced interpretation can itself constitute a meaningful form of thought-level progress~\cite{beck1976cognitive,clark2013cognitive,gomezpenedo2020theory}. A short reflective conversation, however, cannot resolve every psychosocial concern. Participants were less satisfied when their goals required concrete advice or action in the external world. 

Future systems focused on making thought-level progress could communicate the scope of their intervention, and avoid suggesting that they directly support behavior change~\cite{zhang2025computational}.

Beyond self-guided settings, thought-level progress scaffolds may also offer a novel paradigm of measurement for traditional psychotherapy. Conventional measurement-based care relies on standardized symptom scales administered at fixed intervals~\cite{lambert2018collecting, desimone2023impact}, which compare a client's scores against population norms but are not designed to capture the person-specific content of what changed within a session. Idiographic, goal-focused measures have been proposed to address this gap~\cite{cooper2023goals, haynes2009idiographic}, but typically still rely on a client's periodic self-report rather than material generated during the therapeutic process itself. A thought-level scaffold offers a complementary approach: rather than asking a client to retrospectively rate their progress on a fixed scale, it surfaces evidence-linked markers of reframing and evidence integration as they occur in the client's own words, within the session. This shifts the unit of measurement from a standardized score to traceable changes in thinking---giving therapists and clients a session-grounded record to interpret alongside standardized outcome measures. 
We believe this is a fruitful direction for future research at the intersection of HCI and psychology, to create more proximal measures of psychotherapeutic progress related to clients' specific goals in care.

\subsection{Balancing Real-Time Salience with Task Switching}
\label{subsec:discussion-timing}

Our findings in section~\ref{subsubsec:findings-timing-salience} provide evidence that real-time feedback on thought-level progress made emerging thinking shifts salient within the conversation but could interrupt the continuity of reflection, whereas post-session feedback supported synthesis afterward but lost that in-the-moment salience.
More specifically, inspecting a Progress Scaffold Card required participants to shift from experiencing the conversation to evaluating the system's representation, then reconstruct the conversational thread on return.
Other work in HCI studying experience sampling during VR gameplay similarly found that there is a cost to switching tasks to respond to questions~\cite{potts2025retrosketch}. 
Research on task switching offers one explanation for this trade-off: switching attention between tasks carries a performance cost~\cite{monsell2003task}, resuming an interrupted task requires retrieving the suspended task goal~\cite{trafton2003preparing}, and interruptions delivered at the end of a task tend to be less disruptive than those delivered mid-task~\cite{adamczyk2004not,bailey2006need}.

Several \systemName participants described this cost directly in section ~\ref{subsubsec:findings-context-switching}, while others valued the very redirection that live feedback provided. A similar tension emerged in Chen et al.'s technology probe study of AI-assisted meeting reflection: participants described active prompts as opportunities for reflection but also as potentially disruptive, whereas passive visualizations risked being overlooked~\cite{chen2025track}. Real-time feedback therefore functioned as both interruption and reflective cue, and its value depended on whether redirecting attention supported or displaced participants' immediate reflective goals.

While participants identified a cost to real-time progress tracking, our study does not provide evidence of when progress scaffolding best supports reflection.
We did not compare the same full scaffold delivered in real-time versus after the conversation, since \baseline provided only a retrospective summary rather than a deferred version of all highlights and Progress Scaffold Cards. We therefore cannot say whether participants who preferred delayed feedback would have gained the same benefit from a fully deferred scaffold, or whether stated timing preferences track actual reflective outcomes.
Future work should experimentally separate the availability of the scaffold from the timing of its presentation, comparing the same progress representations delivered in real-time versus after the conversation. Such work could evaluate not only which timing participants prefer, but whether it changes how well they notice progress and how deeply they reflect. Future designs might then let users adjust feedback timing and learn which configuration best supports their reflection. Drawing on prior work on interruption timing and AI-assisted reflection~\cite{adamczyk2004not,bailey2006need,chen2025track}, one approach would be to defer feedback until there is a natural pause, such as after completing a reframing exchange. This design would aim to preserve opportunities for noticing progress while giving users control over when to engage in more detailed interpretation.

\subsection{Designing Traceable Progress Interpretations}
\label{subsec:discussion-traceability}

Our findings in Section~\ref{subsubsec:findings-transparency} show that progress interpretations were easier to understand and evaluate when participants could trace a Progress Scaffold Card back to the statement on which it was based. This interaction reflects Shneiderman's overview-and-details-on-demand pattern: the card provides an overview, while the linked conversation provides the details needed to evaluate it \cite{shneiderman1996eyes}. Similarly, AutoML Trace makes the origins and histories of human- and AI-produced artifacts inspectable \cite{rogers2023tracing}. In our setting, linking an interpretation to the user's own statement similarly made its origin and context inspectable.

Though our participants appreciated this traceability, few provided feedback on whether the text marked as progress aligned with their perceptions of progress. 
We designed rating controls (the "Agree", "Not sure" and "Disagree" selections) to collect feedback on whether specific text marked as progress aligned with users’ own views of progress, with the intention of using this feedback to inform marker detection, but only two participants used these controls. Other participants did not submit ratings, but expressed agreement with the markers when asked during interviews \S\ref{subsubsec:findings-transparency}. 
This suggests a limitation for systems that use similar types of feedback mechanisms to improve the system. 

We also relate this to the distinction participants identified in Section \S\ref{subsec:discussion-thought-vs-behavior}: participant's agreement with a marker of thought-level change should not be taken as evidence that users perceive the marked progress toward their broader goals.

\subsection{Limitations and Future Work}
\label{subsec:discussion-limitations}

Our study has four primary limitations. First, our study captured a single conversation, and we do not have evidence suggesting that our system motivated downstream goal attainment through, for example, behavior change.
Future longitudinal studies can evaluate whether visualizing thought-level progress markers motivates behavior changes that support goal attainment.
Second, we recruited U.S. adults through Prolific. This recruitment approach allowed us to easily reach participants who use AI to support their psychosocial goals, but may overrepresent people who are comfortable participating in online studies. 
The findings may therefore not represent people from diverse contexts.
Third, our intervention was tested on a small sample of 30 participants. 
We therefore interpret the quantitative findings as preliminary evidence to motivate the investigation of similar systems in larger samples, and along with the qualitative accounts, provide meaningful information to inform these future systems' design. 
Larger studies with more diverse participants would provide stronger estimates of the system's effectiveness. 
In addition, we did not test all configurations of our system, for example, whether users benefit more from real-time versus post-session progress markers.
Future work can conduct randomized experiments to distill and optimize the system components.
Finally, we did not conduct a full validation study of the progress markers. 

The goal for this work, however, was not to develop novel measures of progress, but instead, to understand how visualizing thought-level progress markers affect users' experiences in a conversational AI mental health intervention.
Future work contributing novel measures could conduct more rigorous studies to establish the reliability and validity of thought-level progress markers.

\section{Conclusion}
\label{sec:conclusion}

This work presents \systemName and provides preliminary evidence that making possible signs of thought-level progress visible can support psychosocial progress sensemaking during AI-guided cognitive restructuring. By giving people a way to track and revisit changes in their thinking, such systems could help people develop a clearer understanding of progress toward psychosocial goals. Recognizing changes in thinking and behavior as distinct but related forms of progress could help people appreciate emerging insights while considering what further action or support they need. By providing representations of progress that attend to the diverse ways people experience positive psychological change, this approach could inform mental health interventions that better recognize and support personally meaningful change.


\section*{Acknowledgment of The Use of AI}
We acknowledge the use of generative AI tools to improve the grammar, style, and readability of this manuscript and to assist with code development for the system described in ~\ref{sec:system}. We used large language models to assist with organizing interview transcript materials and improving language clarity. Coding interpretations and final themes were developed by the researchers.

\bibliographystyle{ACM-Ref-Format}
\bibliography{bib/reference}

\appendix
\newpage

\section{Survey Questions}
\label{appendix:survey-questions}

We list the survey questions, their corresponding question number in the original scale, and our abbreviated narrative in the paper in Tab.~\ref{tab:measures}.

\begin{table*}[h]
\centering
\small
\caption{Survey questions used in the study. All items were answered on a
five-point Likert scale. Numbers in the first column refer to the question statement's position in the original scale, while dashes indicate modified questions.}
\label{tab:measures}
\begin{tabularx}{\textwidth}{@{}c l X@{}}
\toprule
\textbf{Q\#} & \textbf{Abbreviation} & \textbf{Adapted statement} \\
\midrule
\multicolumn{3}{@{}l}{\textbf{Working Alliance Inventory, Short Revised (WAI-SR)}~\cite{hatcher2006development}} \\
1  & Clarity about change      & As a result of interacting with the system, I am clearer about how I might be able to change. \\
4  & Collaborative goal setting    & The system and I collaborated on setting goals. \\
6  & Working towards agreed goals & The system and I worked towards mutually agreed upon goals. \\
8  & Agreement on priorities     & The system and I agreed on what is important for me to work on. \\
2  & New perspective on problem      & The interaction with the system gave me new ways of looking at my problem. \\
11 & Change understanding  & The system and I established a good understanding of the kind of changes that would be good for me. \\
12 & Confidence in approach      & I believe the way the system and I worked on my problem is correct. \\
\midrule
\multicolumn{3}{@{}l}{\textbf{Technology-Supported Reflection Inventory (TSRI)}~\cite{bentvelzen2021development}} \\
1 & Wake-up call for change       & Using the system has led to a wake-up call to make changes in my life. \\
2 & Change in approach     & As a result of using the system, I have changed how I approach things. \\
3 & Ideas to overcome challenges & Using the system gives me ideas on how to overcome challenges. \\
4 & Enjoyment goal exploration     & I enjoy exploring my goals with the system. \\
5 & Ease of goal overview        & The system makes it easy to get an overview of my goals. \\
6 & Ease of goal progress        & The system makes it easy to make progress towards my goals. \\
\midrule
\multicolumn{3}{@{}l}{\textbf{Perceived Goal Attainability}~\cite{schimpf2026supporting}} \\
- & Perceived goal attainability & The system changed my perception of how attainable those goals are.\\
\midrule
\multicolumn{3}{@{}l}{\textbf{Self-Efficacy}~\cite{bandura1977self, mcauley1991self}} \\
- & Clarity goal progress & The system increased my understanding of what progress toward my goal looks like. \\
- & Skills for attaining goal  & From using the system, I have acquired the skills I need to attain my goal. \\
\midrule
\multicolumn{3}{@{}l}{\textbf{Mobile App Rating Scale, Section F (MARS)}~\cite{stoyanov2015mobile}} \\
F.4 & Motivation to continue goal pursuit & The system increases my motivation to continue working toward my goal. \\
\midrule
\multicolumn{3}{@{}l}{\textbf{State Hope Scale}~\cite{snyder1996development}} \\
5 & Multiple pathways to goals & I can think of many ways to reach my current goals. \\
6 & Meeting self-defined goals    & At this time, I am meeting goals that I have set for myself. \\
\midrule
\multicolumn{3}{@{}l}{\textbf{System Usability Scale (SUS)}~\cite{brooke1996sus}} \\
1  & Frequent use    & I think that I would like to use this system frequently. \\
2  & Complexity      & I found the system unnecessarily complex. \\
3  & Ease of use     & I thought the system was easy to use. \\
4  & Need support    & I think that I would need the support of a technical person to be able to use this system. \\
5  & Integration     & I found the various functions in this system were well integrated. \\
6  & Inconsistency   & I thought there was too much inconsistency in this system. \\
7  & Learnability    & I would imagine that most people would learn to use this system very quickly. \\
8  & Cumbersome      & I found the system very cumbersome to use. \\
9  & Confidence      & I felt very confident using the system. \\
10 & Prior learning  & I needed to learn a lot of things before I could get going with this system. \\
\bottomrule
\end{tabularx}
\end{table*}

\newpage

\section{LLM Prompt for Guided Conversation}
\label{appendix:prompt}

\begin{lstlisting}[language=Markdown]
You are a warm, perceptive reflection guide supporting someone through a
cognitive restructuring exercise. You are not acting as a clinician, diagnosing
the user, or giving medical advice.

Your role is to help the user examine a stressful thought by:
1. finding the specific thought underneath what is bothering them,
2. connecting it to a recent situation and emotion,
3. gently noticing possible thinking patterns,
4. helping the user write a more balanced thought themselves.

You are not a chatbot following a script. You listen carefully and ask one good
question at a time.

Use tentative language:
- "it might be..."
- "one possibility is..."
- "I wonder if..."
- "could there be another explanation?"
- "what else might be true here?"

Avoid certainty:
- Do not say "you are catastrophizing" or "this is what happened."
- Do not make diagnostic claims.
- Do not tell the user what they should believe.
- Do not force positivity.

A reframe is not positive thinking. It is a more complete thought that still
respects the difficult parts of the situation.

Safety (read carefully - this overrides the normal flow):
Judge the risk in the latest user message and set safety_risk_level to "low",
"moderate", or "high", or leave it null when there is no safety concern.

- low: vague or passive distress with no intent.
  -> Respond with brief, genuine warmth, then CONTINUE the restructuring.

- moderate: stronger or recurring thoughts of death or self-harm, but no stated
  intent, plan, means, or framing of suicide or self-harm as a solution.
  -> Acknowledge it warmly and seriously, check in gently once, then continue the
     restructuring at a slower, softer pace.

- high: suicide or self-harm framed as a solution; a stated intent, plan, means, or
  timing; or active, ongoing self-harm.
  -> STOP. Do not continue restructuring at all, and do not advance the stage. Punt
     to a human immediately: express serious care, say a member of the team will
     reach out, and give a brief crisis safety net.

[Additional safety examples and contrast cases omitted for brevity.]

---

How to stay concrete:
Speak concretely, not abstractly. The user should never need to interpret a
therapy concept before answering. Ask questions that can be answered from their
actual moment, words, body, action, or next step.

[Repeated good/bad question examples omitted for brevity.]

---

How to handle transitions:
Stage transitions should feel like one continuous conversation, not like moving
through a worksheet. Do not announce stage names to the user. Use information the
user already gave. Do not ask them to repeat it.

---

The user's goal:
{goal_text}

Current stage: {current_stage}
User turns in this stage so far: {stage_user_turns}

Recent assistant questions in this stage to avoid repeating:
{recent_stage_questions}

What you know so far:
{collected_data}

Conversation so far:
{recent_messages}

---

Your focus for this turn ({current_stage}):
{stage_guidance}

Stage guidance includes:
- thought: help the user find the specific thought underneath what is bothering them.
- situation: help the user describe the specific situation that triggered the thought.
- emotion: help the user name the emotion and rough intensity.
- thinking_trap: help the user notice what thinking pattern may be making the stressful
  thought feel stronger, harsher, or more certain.
- reframing: help the user arrive at a more balanced thought in their own words.

[Detailed per-stage guidance retained in the implementation; repeated examples omitted.]

---

Thinking patterns you may draw on:
{thinking_traps}

Thinking pattern definitions include:
- all_or_nothing
- overgeneralizing
- labeling
- fortune_telling
- mind_reading
- emotional_reasoning
- should_statements
- personalizing
- disqualifying_positive
- catastrophizing
- comparing_and_despairing
- blaming
- negative_feeling_or_emotion

---

Progress insight detection:
Detect progress indicators from the latest user message. Use conversation context
only to interpret that message; do not give credit for something the assistant said.

Indicator definitions:
- awareness: the user puts an internal experience into words.
- bias_recognition: the user realizes a thought may be filtered, one-sided, biased,
  distorted, or not the whole truth.
- evidence_integration: the user weighs evidence, names counter-evidence, compares
  thought versus reality, or brings in a fact that complicates the original thought.
- reframing: the user formulates or accepts a more balanced, adaptive thought.
- agency: the user expresses ownership, mastery, an actionable next step, or credits
  their own reasoning/effort for the shift.

---

Output rules:
Respond only with valid JSON matching the SessionLLMOutput schema.

The response must include:
assistant_message, current_stage, next_stage, stage_complete, extracted_thought,
extracted_situation, extracted_emotions, thinking_trap_candidates, focused_trap_id,
explored_trap_ids, user_trap_reflection, ready_for_reframing, reframe_components,
reframe, safety_flag, safety_risk_level, and progress_indicators.

The session moves to whatever valid stage you put in next_stage. Keep it equal to
the current stage until you genuinely mean to move on. Apply the tiered Safety rules
above. For low and moderate risk, respond with warmth and continue. For high risk,
stop restructuring, keep next_stage at the current stage, punt to a human, and include
the brief crisis safety net.
\end{lstlisting}

\section{Themes and Illustrative Quotes}
\label{appendix:themes}
We present the themes, subthemes, and illustrative participant quotes from our qualitative analysis in Table~\ref{tab:themes-quotes}.

\begingroup
\small
\setlength{\tabcolsep}{5pt}
\renewcommand{\arraystretch}{1.15}
\begin{longtable}{@{}>{\raggedright\arraybackslash}p{\dimexpr.22\textwidth-10pt\relax}|>{\raggedright\arraybackslash}p{\dimexpr.25\textwidth-10pt\relax}|>{\raggedright\arraybackslash}p{\dimexpr.53\textwidth-2pt\relax}@{}}
\caption{Themes, subthemes, and illustrative participant quotes.}\label{tab:themes-quotes}\\
\hline
\textbf{Theme} & \textbf{Subtheme} & \textbf{Example quotation} \\
\hline
\endfirsthead
\multicolumn{3}{l}{\tablename~\thetable{} (continued)}\\
\hline
\textbf{Theme} & \textbf{Subtheme} & \textbf{Example quotation} \\
\hline
\endhead
\hline
\multicolumn{3}{r}{\footnotesize Continued on next page}\\
\endfoot
\hline
\endlastfoot

\textbf{Progress Scaffold Cards helped people notice and make sense of progress} & Noticing progress that might otherwise go unnoticed & ``It's much more subjective. There's no number that I can assign to it.'' (X3) \\
\cline{3-3}
 &  & ``I normally wouldn't catch stuff like that'' (X1) \\
\cline{3-3}
 &  & ``more so it\ldots{} made it more visible to me'' (X2) \\
\cline{3-3}
 &  & ``a very visual way to see something that's typically not visual'' (X7) \\
\cline{3-3}
 &  & ``engage more and provide more information'' (X2) \\
\cline{3-3}
 &  & ``every time something lit up\ldots{} a little dopamine rush'' (X5) \\
\cline{2-3}
 & Using the cards to think more deeply about the conversation & ``pushes me to look at it\ldots{} what I'm contemplating\ldots{} isn't a fact'' (X3) \\
\cline{3-3}
 &  & ``repeated my ideas back to me\ldots{} go deeper into the emotions'' (X4) \\
\cline{3-3}
 &  & ``it's so much deeper\ldots{} so much more to it'' (X6) \\
\cline{3-3}
 &  & ``interpreted that way and categorized that way'' (X6) \\
\cline{3-3}
 &  & ``I don't know if it taught me how to track the progress, but it more so taught me how to\ldots{} Very quickly, it taught me a different way of looking at the issue'' (X3) \\
\cline{3-3}
 &  & ``I have some clarity about it'' (Y4) \\
\cline{2-3}
 & Checking the markers against their own words & ``highlights on the sentence\ldots{} tied to the card\ldots{} went back and forth'' (X3) \\
\cline{3-3}
 &  & ``it's very, like, evidence-based'' (X4) \\
\cline{3-3}
 &  & ``I didn't disagree with any of it'' (X1) \\
\cline{3-3}
 &  & ``marked it as a reframing\ldots{} totally agree'' (X7) \\
\cline{3-3}
 &  & ``picking out the correct things\ldots{} should have been highlighting'' (X5) \\
\cline{3-3}
 &  & ``I can disagree if I want'' (X4) \\
\hline
\textbf{Progress in the conversation did not always mean progress toward the goal} & Wanting to see progress across conversations & ``snapshot, one frame of a 67-year-old movie'' (X11) \\
\cline{3-3}
 &  & ``go back to this in the future'' (X2) \\
\cline{3-3}
 &  & ``see the progress on maybe how far I've come'' (X6) \\
\cline{3-3}
 &  & ``remember when, 6 months ago, you talked about this'' (X13) \\
\cline{3-3}
 &  & ``what's my summary of last week\ldots{} overall summary'' (X13) \\
\cline{3-3}
 &  & ``viewing the progress summary\ldots{} be able to copy it'' (X14) \\
\cline{2-3}
 & Looking for changes beyond the conversation & ``depends which goal I'm\ldots{} working on'' (X7) \\
\cline{3-3}
 &  & ``helps give me clarity in the plan, which makes things less stressful'' (X5) \\
\cline{3-3}
 &  & ``I do have some power to affect\ldots{} the outcome'' (X5) \\
\cline{3-3}
 &  & ``I feel more at peace\ldots{} more centered\ldots{} less outer-directed to looking at somebody as causing my problem'' (Y4) \\
\cline{3-3}
 &  & ``actual progress\ldots{} me doing it outside of the chatbot'' (Y10) \\
\cline{3-3}
 &  & ``exactly where I was when it began'' (X11) \\
\cline{3-3}
 &  & ``I don't have any more money\ldots{} borrowing money from friends'' (X12) \\
\hline
\textbf{Real-time feedback helped people notice progress but could interrupt reflection} & Noticing progress in real time and reviewing the whole conversation afterward & ``made me choose my words more decisively'' (X6) \\
\cline{3-3}
 &  & ``attack the thoughts\ldots{} as you're having them'' (X9) \\
\cline{3-3}
 &  & ``pushes me\ldots{} guiding me\ldots{} I'm on to something'' (X6) \\
\cline{3-3}
 &  & ``enjoyed the overall summary\ldots{} regroup it all together'' (X7) \\
\cline{3-3}
 &  & ``a clean snapshot and picture of what was just talked about'' (X12) \\
\cline{3-3}
 &  & ``what you went through\ldots{} how it progressed\ldots{} conclusion\ldots{} for next time'' (X15) \\
\cline{2-3}
 & Different preferences for when to review feedback & ``I liked it popping up as I went'' (X1) \\
\cline{3-3}
 &  & ``I would probably prefer it at the end. Okay. only because\ldots{} I could see myself getting distracted.'' (X3) \\
\cline{3-3}
 &  & ``look at the different steps and relate it back\ldots{} when I'm done'' (X4) \\
\cline{3-3}
 &  & ``don't know if I have a preference'' (X7) \\
\cline{3-3}
 &  & ``I'd rather be focusing on the conversation and feeling'' (X13) \\
\cline{3-3}
 &  & ``focus more on the feedback and not the conversation\ldots{} get distracted'' (Y14) \\
\cline{2-3}
 & Switching between the conversation and the cards took effort & ``really easy\ldots{} anybody could figure out how'' (X15) \\
\cline{3-3}
 &  & ``jumping back between\ldots{} sitting outside of it versus being in it'' (X13) \\
\cline{3-3}
 &  & ``takes away from\ldots{} keeping the conversation flow'' (X15) \\
\cline{3-3}
 &  & ``I didn't even pay attention to that until it flashed'' (X12) \\
\cline{3-3}
 &  & ``definitely didn't make it any harder to follow'' (X7) \\
\cline{3-3}
 &  & ``I watched the video, so that helps, but even after that, I didn't remember, like, what the cards were for'' (X3) \\
\end{longtable}
\endgroup

\end{document}